\documentclass[sn-apa]{sn-jnl}

\usepackage{graphicx}%
\usepackage{hyperref}
\usepackage{makecell}
\usepackage{multirow,bm,bbm}%
\usepackage{amsmath,amssymb,amsfonts}%
\usepackage{amsthm}%
\usepackage{mathrsfs}%
\usepackage[title]{appendix}%
\usepackage{xcolor}%
\usepackage{textcomp}%
\usepackage{manyfoot}%
\usepackage{booktabs}%
\usepackage{algorithm}%
\usepackage{algorithmicx}%
\usepackage{algpseudocode}%
\usepackage{listings}%
\usepackage{subcaption}
\usepackage{booktabs}
\usepackage{amsmath}
\usepackage{tikz}
\usetikzlibrary{bayesnet}
\usepackage{booktabs}      
\usepackage{longtable}     
\usepackage{array}         
\usepackage{multirow}      
\usepackage{multicol}      
\usepackage{caption}       

\usepackage{threeparttable}
\usepackage[table]{xcolor} 
\usepackage{makecell}      

\usepackage{verbatim}

\theoremstyle{thmstyleone}%

\theoremstyle{thmstyletwo}%

\theoremstyle{thmstylethree}%

\newcount\Comments  
\usepackage{color}
\definecolor{darkgreen}{rgb}{0,0.5,0}
\definecolor{purple}{rgb}{1,0,1}
\newcommand{\kibitz}[2]{\ifnum\Comments=1\textcolor{#1}{#2}\fi}

\begin{document}

\title[]{A comparison between initialization strategies for the infinite hidden Markov model}

\author*[1,2]{\fnm{Federico P.} \sur{Cortese}}\email{federico.cortese@unimi.it}

\author[1,3]{\fnm{Luca} \sur{Rossini}}\email{luca.rossini@unimi.it}
\affil*[1]{\orgdiv{Department of Economics, Management, and Quantitative Methods}, \orgname{University of Milan}, \orgaddress{\street{Via Conservatorio 7}, \city{Milan}, \postcode{20122}, \country{Italy}}}
\affil[2]{\orgdiv{Institute for Applied Mathematics and Information Technologies},
\orgname{National Research Council}, \orgaddress{\street{Via Alfonso Corti, 12}, \city{Milan}, \postcode{20133}, \country{Italy}}}
\affil[3]{\orgdiv{Fondazione Eni Enrico Mattei}, 
\orgaddress{\street{Corso Magenta 63}, \city{Milan}, \postcode{20123}, \country{Italy}}}

\abstract{

\textbf{Abstract.} 
Infinite hidden Markov models provide a flexible framework for modeling time-series with structural changes and complex dynamics, without requiring the number of latent states to be specified in advance. This flexibility is achieved through the hierarchical Dirichlet process prior, while efficient Bayesian inference is enabled by the beam sampler, which combines dynamic programming with slice sampling to truncate the infinite state space adaptively.
Despite extensive methodological developments, the role of initialization in this framework has received limited attention. This gap is addressed by systematically evaluating initialization strategies commonly used for finite hidden Markov models and assessing their suitability in the infinite setting.
Results from both simulated and real datasets show that distance-based clustering initializations consistently outperform model-based and uniform alternatives, the latter being the most widely adopted in the existing literature.

}


\maketitle

\section{Introduction}
\label{sec:intro}

Regime-switching models offer a flexible framework for modeling multivariate time-series characterized by structural changes or dynamic behaviors over time. 
These models assume the existence of an unobserved latent process that switches between different (usually called $K$) regimes.
These shifts among regimes allow for capturing nonlinearities, sudden breaks, and shifts in dynamics that standard models often fail to represent adequately.
Based on their flexibility, regime-switching models have been extensively applied in a wide range of applications, including signal processing \citep{rabiner2002tutorial}, energy forecasting \citep{erlwein2010hmm}, 
and finance \citep{nystrup2018dynamic}.  

Within this family, hidden Markov models \citep[HMM,][]{bart:farc:penn:13,zucchini:2017} are particularly effective since they combine latent state dynamics with 
state-dependent
emission distributions that can be specified by the user.  
However, a long-standing challenge in the use of finite HMMs concerns model selection \citep{Buckby2020}, 
particularly determining the appropriate number of hidden states $K$. 
Since this number is not known a priori, it must be selected through different methods. Common approaches include cross-validation \citep[CV,][]{stone1974cross}, evaluating out-of-sample performance across candidate values, or the use of penalized likelihood (or information) criteria, such as the Akaike’s information criterion \citep[AIC,][]{akaike1974new}, the Bayesian information criterion \citep[BIC,][]{schwarz1978estimating}, and the integrated completed likelihood \citep[ICL,][]{biernacki2002assessing}. 
Although these methods are effective in real data applications, they can be computationally demanding,  
as they require fitting models over a grid of hyperparameters.
Moreover, they can also yield results that depend on sample size, initialization, and choice of the information criteria (see, for example, \cite{bart:farc:penn:13} and \cite{cortese2024maximum}).

To overcome these issues, Bayesian nonparametric models have emerged as a natural alternative. 
These models allow the data to determine the model complexity by assuming an infinite number of latent components, while only a finite subset is effectively used. 
The foundational idea stems from infinite mixture models \citep{escobar1995bayesian}, which employ a Dirichlet process \citep[DP,][]{ferguson1973bayesian} prior over the component distributions. The DP prior allows clustering among data points and it can be extended to hierarchical settings through the hierarchical Dirichlet process \citep[HDP,][]{teh2006hierarchical}, which allows multiple mixture models to share a common set of components.

Within the Bayesian nonparametric framework, the infinite hidden Markov model \citep[iHMM,][]{beal2001infinite} represents a seminal contribution. 
This model replaces the transition matrix of the latent Markov process with a HDP prior on it, allowing for a countably infinite set of latent regimes. 
Its formulation allows the model to automatically infer the number of regimes without the need for explicit model selection. Moreover, the iHMM can be viewed as a fully Bayesian nonparametric extension of the HMM where both the transition probabilities and the number of regimes are countably infinite and are inferred directly from the data, which determines the model complexity.

Since its introduction, the iHMM has been widely extended and applied to different applications. 
For instance, 
\cite{song2014modelling} specifies an iHMM with Gaussian autoregressive (AR) regimes for U.S. real interest rates, while 
\cite{shi2015identifying} use an iHMM to detect speculative bubbles in U.S. housing ratios. \cite{maheu2016infinite} incorporate the iHMM within a continuous time Vasicek model for U.S. short-term interest rates, and \cite{hou2017infinite} applies a vector autoregressive (VAR) iHMM to macroeconomic forecasting. 
\cite{yang2019stock} extends the iHMM to capture relationships between stock returns and economic growth, while \cite{jin2016bayesian} introduce an iHMM with Wishart observations for low-dimensional realized covariance matrices, and \cite{jin2019bayesian} extend it to high-dimensional realized covariance matrices via dynamic factor models, applied to large S\&P 500 panels.
Beyond financial econometrics, 
\cite{hoskovec2023infinite} develop an iHMM for asynchronous multivariate series with missing data using a covariate-dependent probit stick-breaking process, applied to air pollution exposure.

Despite the growing interest in iHMMs, an important aspect concerns the role of {initialization} in posterior inference. 
Estimation of the iHMM typically relies on the beam sampler \citep{van2008beam}, a blocked Gibbs sampling algorithm that efficiently handles the infinite latent state space. 
Although \citet{van2008beam} argue that the beam sampler is robust to initialization and recommend using a uniform initialization for the latent states, no systematic analysis has yet examined how different initialization strategies affect its behavior.
In contrast, the literature on finite HMMs and mixture models shows that inference results, particularly when relying on the expectation–maximization (EM) algorithm \citep{dempster:1977}, can strongly depend on the initialization of latent states 
\citep{pandolfi2021maximum}.
For finite HMMs, \citet{maruotti2021initialization} 
perform an extensive comparison of several initialization strategies, such as $k$-means, partitioning around medoids \citep{kaufman1987clustering,kaufman2009finding}, hereafter referred to as \textit{pam}, and \textit{Gaussian mixtures} \citep{mclachlan2000finite}.
Their results indicate that $k$-means often performs best overall, Gaussian mixtures are more effective in high-dimensional settings, and pam offers advantages in low-dimensional cases.
However, these results are limited to finite-state HMMs and to data generated from a multivariate Gaussian distribution, leaving open the question of how initialization affects inference in nonparametric, infinite-state models.

Our main purpose is to examine how different initialization strategies influence the performance of the beam sampler in the multivariate Gaussian iHMM. 
To the best of our knowledge, this is the first study to systematically investigate initialization strategies within the iHMM framework. 
Indeed, we aim to bridge the gap between finite HMMs and iHMMs by evaluating how initialization methods commonly employed in the former model perform when applied to the latter. Specifically, we consider four standard approaches used in classical HMM estimation: uniform initialization as proposed by \cite{van2008beam}, $k$-means and pam based on the GAP statistic \citep{witten2010framework}, and Gaussian mixture models \citep{mclachlan2000finite} guided by information criteria such as the BIC.

To test these methods, we perform two extensive simulation studies: one considering Gaussian data and one with Student-$t$ data with fat tails, to assess performance under both standard and heavy-tailed conditions, the latter being a setting that better reflects real-world scenarios. 
This is done to investigate the performance of each initialization strategy under a misspecified model formulation, and consequently to assess the robustness of both the beam sampler and the initialization strategies.
We evaluate the performance of the strategies by testing (i) their ability to recover the true latent structure, measured by the adjusted Rand index (ARI; \citealp{hubert:1985}) at convergence, and (ii) their convergence speed, measured by changes in ARI across iterations.
As a further measure, we evaluate the ability of the beam sampler and the initialization strategies to converge through the \citet{Geweke1992} diagnostic and the autocorrelation time (ACT).

Once we assess the performance of the initialization methods in simulation, we apply the iHMM to two real datasets: (1) the monthly industrial production volume index for nine European countries from March 2002 to August 2025, and {(2) daily financial market data comprising European equity and bond indices, gold, Bitcoin, the EUR--USD exchange rate, and Brent crude oil prices for the period from January 2019 to March 2026.}

Results indicate that $k$-means and pam generally outperform Gaussian mixture initializations, particularly when the data-generating process violates the Gaussianity assumption or when different clusters exhibit substantial overlap. In contrast, the uniform initialization performs poorly in nearly all scenarios, suggesting that it should be avoided in practice.
The choice of initialization also affects model performance on real data. 
For both real data applications, initialization has a substantial impact on the final model estimates, with $k$-means and pam yielding consistent and more robust classification.

The paper is organized as follows. Section \ref{sec:iHMM} presents the formulation of the multivariate Gaussian iHMM and the beam sampler. Section \ref{sec:init} discusses the initialization strategies considered, while Section \ref{sec:simstud} shows two extensive simulation studies assessing their accuracy and convergence under different scenarios.
Section~\ref{sec:empstud} presents two empirical studies where we model real-world time-series with an iHMM.
Section \ref{sec:conclusions} concludes.

\section{Model formulation and posterior inference}
\label{sec:iHMM}

Selecting the appropriate number of components in mixture and regime-switching models is a well-known challenge. 
Classical approaches typically rely on likelihood-based model selection criteria or cross-validation, which are often sensitive to the number of regimes specified a priori.
Bayesian nonparametric methods offer an elegant alternative by allowing the model complexity to adapt flexibly to the data. 
Rather than fixing the number of latent states in advance, these models infer it directly from the data through prior constructions that support a potentially infinite number of components
\citep{ghosh2003bayesian, hjort2010bayesian, muller2015bayesian}. 

This approach is based on the Dirichlet process \citep[DP,][]{ferguson1973bayesian}, which defines a prior over distributions and induces mixture models with 
a countably infinite
number of components. 
Several extensions of the DP mixtures have been proposed, demonstrating its versatility in classification and density estimation \citep{fuentes2010probability, gutierrez2014bayesian}. The HDP introduced by \cite{teh2006hierarchical} generalizes this framework by allowing multiple groups to share a common set of mixture components. 
This hierarchical construction provides the foundation for the iHMM.
\vspace{2em}

\subsection{Infinite hidden Markov model}
\label{subsec:iHMM}

We begin by briefly revisiting the finite HMM before introducing its infinite, nonparametric extension. 
Consider a multivariate 
time-series $\textbf{y}_1,\ldots, \textbf{y}_T$, where $\textbf{y}_t \in \mathbb{R}^P$,  
and assume the existence of a latent first-order Markov process $\mathbf{s}$ with components
\begin{align*}
   s_1, \dots, s_T, \qquad s_t \in \mathbb{N}, 
\end{align*}
where each $s_t$ takes values in $\{1,2,\ldots,K\}$. 
The model for $s_1,\ldots,s_T$ specifies transition probabilities $\pi_{ij}(t)=\mathbb{P}(s_t=j|s_{t-1}=i)$, and initial state probabilities $\pi_{i}=\mathbb{P}(s_1=i)$, for $i,j \in\{1,\ldots,K\}$. 
Conditional on the latent state $s_t$, the observation $\textbf{y}_t$ is generated from a state-specific emission distribution $f(\cdot|\boldsymbol{\theta}_{s_t})$. Assuming conditional independence of the $\textbf{y}_t$ given the states $s_t$, the joint distribution of states and observations is

\begin{align*}
    \label{eq:jointdistr}
f\left(s_1,\ldots,s_T,\textbf{y}_1,\ldots,\textbf{y}_T\right) &= \pi_{s_1} f(\textbf{y}_1|\boldsymbol{\theta}_{s_1})\prod_{t=2}^T \pi_{s_{t-1},s_t}(t)\, f(\textbf{y}_t|\boldsymbol{\theta}_{s_t}).
\end{align*}
While the classical HMM assumes a finite number of states $K$, this can be restrictive in practice.
The infinite HMM relaxes this constraint by allowing the number of regimes to be potentially unbounded, which is achieved by assigning each row of the transition matrix an infinite-dimensional probability vector 
\begin{align*}
    \boldsymbol{\pi}_k = (\pi_{k1},\pi_{k2},\ldots), \quad k=1,2,\ldots. 
\end{align*}
We place a hierarchical DP prior on each row
\begin{align*}
 \boldsymbol{\pi}_k \,|\, G_0, \alpha \;\sim\; \text{DP}(\alpha, G_0), \qquad G_0 \,|\, \gamma \sim \text{DP}(\gamma, H),   
\end{align*}
where $H$ is the base measure and $\alpha$ and $\gamma$ are concentration parameters.\footnote{For more details on the definition of the HDP, we refer the reader to the Supplementary Material.}
This hierarchical structure induces that state-specific transition distributions $\bm{\pi}_k$ share a common set of states, which supports a countably infinite number of regimes \citep{teh2006hierarchical}. The resulting Markov chain is therefore time-homogeneous.

Using the stick-breaking representation \citep{sethuraman1994constructive}, we write $G_0=\sum_{k^\prime=1}^\infty \beta_{k^\prime}\delta_{\boldsymbol{\theta}_{k^\prime}}$ and $G_{k}=\sum_{k^\prime=1}^\infty \pi_{kk^\prime}\delta_{\boldsymbol{\theta}_{k^\prime}}$, 
where the components $\pmb{\beta}=(\beta_{k'})_{k'=1}^{\infty}\sim \operatorname{GEM}(\gamma)$ (Griffiths, Engen and McCloskey) are mutually independent.
{Specifically,
they 
are defined as 
\[
\beta_{k'} = v_{k'} \prod_{l=1}^{k'-1} (1 - v_l),
\quad
v_{k'} \mid \gamma \sim \operatorname{Beta}(1,\gamma),
\quad k' = 1,2,\ldots.
\]
}
Identifying each $G_k$ with both transition probabilities $\pi_{kk^\prime}$ and emission parameters $\boldsymbol{\theta}_k$ allows us to formally define the iHMM as
\begin{equation}
\begin{aligned}
\boldsymbol{\pi}_k &\sim \text{DP}(\alpha, \boldsymbol{\beta}), \\
\boldsymbol{\beta}  &\sim \text{GEM}(\gamma), \\
\boldsymbol{\theta}_k &\sim H, \\
s_t \mid s_{t-1} &\sim \text{Multinomial}(\boldsymbol{\pi}_{s_{t-1}}), \\
\textbf{y}_t \mid s_t &\sim f(\cdot|\boldsymbol{\theta}_{s_t}).
\end{aligned}
\label{eq:iHMM}
\end{equation}

In this paper, we focus on the multivariate Gaussian iHMM, where conditional on the regime indicator $s_t$, the observations follow
\begin{align*}
\textbf{y}_t \,|\, s_t = k, \boldsymbol{\mu}_k, \boldsymbol{\Sigma}_k \;\sim\; \mathcal{N}_P(\boldsymbol{\mu}_k, \boldsymbol{\Sigma}_k),
\end{align*}
and each pair $(\boldsymbol{\mu}_k, \boldsymbol{\Sigma}_k)$ is assigned a Normal-Inverse-Wishart (NIW) prior
\begin{align*}
\boldsymbol{\Sigma}_k \sim \mathcal{IW}(\nu_0, \boldsymbol{\Lambda}_0), 
\qquad
\boldsymbol{\mu}_k \,|\, \boldsymbol{\Sigma}_k \sim \mathcal{N}_P(\boldsymbol{\mu}_0, \boldsymbol{\Sigma}_k/\kappa_0),
\end{align*}
where $\nu_0$ and $\bm{\Lambda}_0$ are the degrees of freedom and the scale matrix, respectively.
Regarding the choice of the hyperparameters, we adopt 
{weakly informative priors, chosen to exert minimal influence on posterior inference while avoiding strict preferences for specific values}.
Specifically, we set $\boldsymbol{\mu}_0 = \bm{0}$, $\boldsymbol{\Lambda}_0 = \boldsymbol{I}_P$, $\kappa_0 = 0.01$, and $\nu_0 = P+2$.

To complete the model specification, we place Gamma hyperpriors on the concentration parameters controlling regime persistence 
{$\alpha \sim \text{Gamma}(a_{\alpha}, b_{\alpha})$, and model complexity $\gamma \sim \text{Gamma}(a_{\gamma}, b_{\gamma})$.}
Following \citet{van2008beam}, we place vague priors by 
setting $a_\alpha=1$, $b_\alpha=1$, $a_\gamma=2$, and $b_\gamma=1$.
This specification {reduces}
the impact on the estimation of the number of states, allowing us to better observe the behavior of each initialization strategy under comparable conditions.

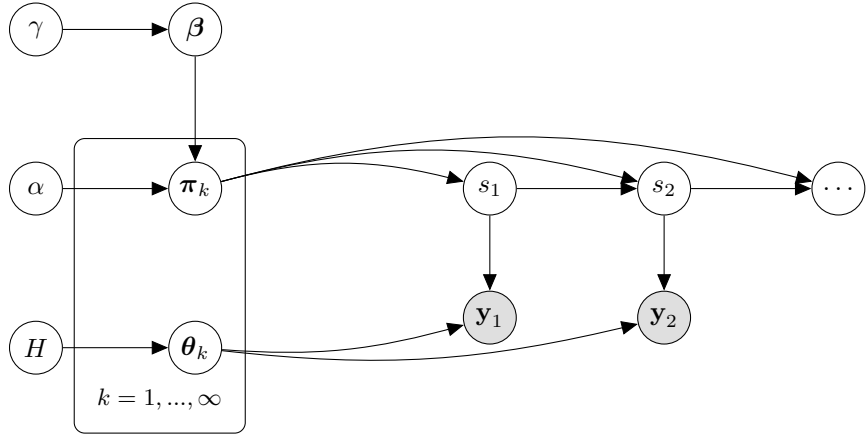
\begin{figure}[ht]
    \centering
\begin{tikzpicture}

\node[latent] (gamma) {$\gamma$};
\node[latent, right=1.4cm of gamma] (beta) {$\boldsymbol{\beta}$};
\node[latent, below=1.4cm of gamma] (alpha) {$\alpha$};
\node[latent, below=1.4cm of beta] (pik) {$\boldsymbol{\pi}_k$};
\node[latent, below=1.4cm of alpha] (H) {$H$};
\node[latent, below=1.4cm of pik] (phik) {$\boldsymbol{\theta}_k$};

\plate [inner sep=0.3cm, xshift=0cm, yshift=0cm] {plateK} {(pik)(phik)} {$k=1,...,\infty$};

\node[latent, right=3.2cm of pik] (s1) {$s_1$};
\node[latent, right=1.6cm of s1] (s2) {$s_2$};
\node[latent, right=1.6cm of s2] (dots) {$\ldots$};

\node[obs, below=1.0cm of s1] (y1) {$\textbf{y}_1$};
\node[obs, below=1.0cm of s2] (y2) {$\textbf{y}_2$};

\draw[->] (gamma) -- (beta);
\draw[->] (beta) -- (pik);
\draw[->] (alpha) -- (pik);
\draw[->] (H) -- (phik);

\draw[->, bend left=15] (pik) to (s1);
\draw[->, bend left=15] (pik) to (s2);
\draw[->, bend left=15] (pik) to (dots);

\draw[->, bend right=10] (phik) to (y1);
\draw[->, bend right=10] (phik) to (y2);

\edge {s1} {s2};
\edge {s2} {dots};
\edge {s1} {y1};
\edge {s2} {y2};

\end{tikzpicture}
\vspace{1.5em}
\caption{Directed Acyclic Graph (DAG) of the infinite hidden Markov model. It exhibits the hierarchical structure of priors, where directed arrows show the causal dependence structure of the model.}
\label{fig:ihmm}
\end{figure}

Figure~\ref{fig:ihmm} provides a graphical representation of the iHMM structure, where the hierarchy between $\bm{\beta}, \bm{\pi}_k$, and $\boldsymbol{\theta}_k$ captures the sharing of regimes across transition rows while allowing an unbounded number of latent states. 

\subsection{The beam sampler}
\label{subsec:beam}

\citet{beal2001infinite} first introduce the iHMM and present a fully Bayesian framework that extends conventional HMMs to a countably infinite number of latent states. 
Their approach relies on DP priors to model the transition and emission distributions, and inference is carried out using an approximate Gibbs sampling scheme that alternates updating the latent state sequence and the model hyperparameters.
 
Building on this work, \citet{teh2006hierarchical}  define the HDP as a general Bayesian nonparametric prior for grouped data, showing that the iHMM arises as a special case of this hierarchical construction. 
This HDP allows each group-specific DP to share atoms through a global base measure drawn from another DP. They also derive an efficient Gibbs sampling algorithm based on the Chinese restaurant franchise and stick-breaking representation.
However, because time-series data typically exhibit strong temporal dependencies between consecutive time steps, Gibbs sampling often converges slowly \citep{scott2002bayesian}. 
To mitigate this, \citet{fox2011sticky} propose an alternative sampling method called the sticky HDP-HMM, which introduces an additional hyperparameter to control self-transition probabilities, while reducing label-switching. 

In this study, we employ the beam sampler introduced by \citet{van2008beam}, due to its computational efficiency and improved convergence properties. 
The method combines \textit{slice sampling} \citep{walker2007sampling, kalli2011slice} with dynamic programming, providing an efficient alternative to Gibbs sampling. 
The key idea is to introduce auxiliary \emph{slice variables} 
{$\boldsymbol{u} = (u_1, \ldots, u_T),$ $u_t
\sim \mathcal{U}(0,\pi_{s_{t-1}s_t}),$} that, at each iteration, restrict the number of feasible state trajectories to a finite subset of the infinite state space, specifically
{$$p(s_t|\textbf{y}_{1:t},u_{1:t}) \propto p(\textbf{y}_t|s_t)\sum_{s_{t-1}:u_t<\pi_{s_{t-1}s_t}} p(s_{t-1}|\textbf{y}_{1:t-1},u_{1:t-1}).$$}
Conditional on these variables, only transitions satisfying $\pi_{s_{t-1}, s_t} > u_t$ remain feasible, thus enabling the use of the forward–filtering backward–sampling (FFBS) algorithm \citep{dempster:1977,chib1996calculating}, a standard tool in finite HMM inference \citep{zucchini:2017}. {Additionally, this augmentation scheme does not change the model, as we can marginalize out $u_t$ and get the original model.}
Subsequently, the transition probabilities $\boldsymbol{\pi}_k$ are drawn from $\text{DP}(\alpha, \boldsymbol{\beta})$, 
with each row of the transition matrix updated according to the current transition counts, 
while the global stick–breaking weights $\boldsymbol{\beta}$ are sampled from a Dirichlet posterior 
using auxiliary table counts, as described in \citet{teh2006hierarchical}.
Emission parameters $\boldsymbol{\theta}_k$ are then resampled given the observations assigned to each state; in the Gaussian case, each pair $\boldsymbol{\theta}_k=(\boldsymbol{\mu}_k, \boldsymbol{\Sigma}_k)$ is updated from a NIW posterior. 
Finally, the hyperparameters $\alpha$ and $\gamma$ are drawn from Gamma posteriors. 

The computational complexity per iteration is $O(TK^2)$ in the worst case, but typically much lower since most transitions are excluded by the slice variables \citep{van2008beam}. 
A schematic implementation of the beam sampler is provided in Algorithm~\ref{alg:beam_sampler}, {while a more detailed description of the sampling algorithm is available in the Supplementary Material}.
\begin{algorithm}[H]
\caption{Beam sampler for the iHMM}
\label{alg:beam_sampler}
\begin{algorithmic}[1]
\State Initialize $\mathbf{s}, \boldsymbol{\pi}_k, \boldsymbol{\beta}, \{\boldsymbol{\theta}_k\}, \alpha, \gamma$
\For{iteration $i= 1, \ldots, N$}
        \State Sample $u_t \sim \mathcal{U}(0, \pi_{s_{t-1}, s_t}),\quad t=1,\ldots,T$
    \State Sample $\mathbf{s}$ via FFBS restricted to $\pi_{s_{t-1}, s_t} > u_t$
    \State Sample $\boldsymbol{\pi}_k \mid \mathbf{s}, \boldsymbol{\beta}, \alpha \sim \text{DP}(\alpha, \boldsymbol{\beta})$
    \State Sample $\boldsymbol{\beta} \mid \mathbf{s}, \gamma$ using auxiliary table counts
    \State Sample emission parameters $\boldsymbol{\theta}_k \mid \mathbf{s}, \textbf{y}_1,\ldots,\textbf{y}_T$ (e.g. NIW posterior)
    \State Sample hyperparameters $\alpha, \gamma$ from Gamma posteriors
\EndFor
\end{algorithmic}
\end{algorithm}

\section{Initialization strategies}
\label{sec:init}

Our main objective is to investigate how different initialization strategies affect the performance of the beam sampler when estimating the multivariate Gaussian iHMM. 
While initialization has been studied extensively in the context of finite HMM literature \citep[see, e.g.,][]{maruotti2021initialization, pandolfi2021maximum}, to our knowledge, our paper is the first work that systematically analyzes initialization procedures in the iHMM framework. 
Because the iHMM generalizes the finite HMM by allowing an unbounded number of regimes, it is not obvious whether initialization methods that work well for finite models will remain effective when the state space is theoretically infinite.

To bridge the gap between finite and infinite hidden Markov models, we evaluate four initialization strategies that are widely used in HMM estimation and clustering analysis. 
These strategies -- uniform initialization, $k$-means clustering, partitioning around medoids (pam), and Gaussian mixture modeling -- represent complementary approaches that differ in their underlying assumptions and computational mechanisms.
By systematically comparing their performance within the iHMM framework, we aim to clarify how initialization choices influence convergence behavior and posterior inference.
\begin{itemize}
    \item[(i)] \textbf{Uniform initialization}. The latent states are assigned randomly as
\begin{align*}
        s_t \sim \mathcal{U}\{1, \dots, K_0\}, \qquad K_0 \sim \mathcal{U}\{2,3,4,5\},
\end{align*}
where $\mathcal{U}\{\ldots\}$ is a discrete uniform distribution.
This procedure, originally proposed by \cite{van2008beam}, provides a baseline where no prior information is imposed. Since the latent allocations are drawn uniformly, it allows the beam sampler to explore the state space broadly initially. 
    \item[(ii)] \textbf{$k$-means clustering}. It partitions the data into $K$ clusters by minimizing the within-cluster sum of squares
\begin{equation*}
\label{eq:wcss}
\min_{\{\boldsymbol{\mu}_k\}_{k=1}^K} \sum_{t=1}^{T} \| \textbf{y}_t - \boldsymbol{\mu}_{s_t} \|^2, \quad s_t \in \{1, \dots, K\},
\end{equation*}
where $\{\boldsymbol{\mu}_k\}_{k=1}^K$ are the centroids. The optimal number of clusters $K$ is determined by using the GAP statistic \citep{witten2010framework}, which compares the observed within-cluster {expected} dispersion to that expected under a null reference distribution
\begin{equation*}
\label{eq:GAP}
\mathrm{GAP}(K) = \mathbb{E}[\log(W_K^*)] - \log(W_K),
\end{equation*}
where \(W_K\) denotes the within-cluster sum of squares for a given partition into \(K\) clusters,
and \(W_K^*\) is the same measure computed from permuted data.  A larger GAP value indicates stronger cluster separation relative to random noise.

We use 25 {random} permutations and compute the statistic through the \textit{clusGap} function from the \textbf{cluster} package \citep{maechler2013package} in \texttt{R} \citep{Rcite}.  
However, since many selection rules can be applied when using the GAP statistic, we adopt the global maximum criterion, which selects the number of clusters $K$ corresponding to the highest GAP value across all candidates. 
This approach is the least conservative\footnote{In the simulation settings, when using $k$-means or pam as initialization strategies, we also applied the \citet{witten2010framework} rule for model selection, which selects the cluster size $k$ as the smallest value satisfying
$
\text{GAP}(k) \geq \text{GAP}(k + 1) - s_{k+1},
$
where $s_{k+1}$ is the standard deviation estimated from the bootstrap samples used to compute $\text{GAP}(k+1)$. 
That said, its tendency to be overly conservative often limited exploration of the state space. 
}, favoring broader exploration of the parameter space. 
    
    \item[(iii)] \textbf{pam algorithm} \citep{kaufman1987clustering,kaufman2009finding}. It is conceptually similar to \(k\)-means but minimizes distances from representative \emph{medoids} instead of centroids. This makes pam more robust to outliers and to non-spherical cluster shapes.
    As for $k$-means, the number of clusters is selected through the GAP statistic with 25 permutations. The model is fitted using the  \textit{pam} function from the \textbf{cluster} \texttt{R} package.
    
    \item[(iv)] \textbf{Gaussian mixture initialization}, hereafter referred to simply as \textit{mixtures}. In this approach, each observation is modeled as arising from a mixture of \(K\) multivariate Gaussian components
    \begin{align*}
           f(\textbf{y}_t) = \sum_{k=1}^{K} \phi_k \, \mathcal{N}_p(\textbf{y}_t \mid \boldsymbol{\mu}_k, \boldsymbol{\Sigma}_k),
    \end{align*}
    where $\phi_k$ are mixture weights satisfying $\sum_k \phi_k = 1$.
    
    The number of components \(K\) is selected according to the Bayesian Information Criterion \citep[BIC,][]{schwarz1978estimating}, defined as
\begin{align*}
    \mathrm{BIC} = -2 \, \ell(\hat{\boldsymbol{\theta}}) + q \, \log(T),
\end{align*}
where \(\ell(\hat{\boldsymbol{\theta}})\) is the maximized log-likelihood, \(q\) is the number of free parameters, and \(T\) is the sample size.  
{\cite{schwarz1978estimating} shows that BIC is consistent and asymptotically selects the correct model order, while \cite{emiliano2014information} provide empirical evidence that BIC outperforms alternative information criteria for moderate and large sample sizes.}
    We fit the model using the \textit{Mclust} function from the \textbf{mclust} \texttt{R} package \citep{fraley2012package}, which automatically fits candidate models and selects the optimal $K$ by BIC comparison.
\end{itemize}

Since GAP statistic and BIC require a maximum number of clusters, we evaluate candidate values of $K$ ranging from 2 to 5, and select the optimal starting point accordingly. 
It is worth noting that this maximum number also constrains the parameter search space; 
however, our choice is justified by both computational reasons and empirical relevance.
From a computational perspective, the GAP statistic requires re-estimating the clustering solution (in our case, via $k$-means or pam) $n$ times, where $n$ is the number of bootstrap samples. This significantly increases computational time, especially in high-dimensional or long time-series settings. 
From an empirical viewpoint, many real-world applications, such as financial time-series segmentation, natural language processing, or gene expression analysis \citep{nystrup:2021}, require a limited number of states for both interpretability and ease of estimation.
Moreover, considering that we place vague hyperpriors on the concentration parameters $\alpha$ and $\gamma$ of the DPs described in Eq. \eqref{eq:iHMM}, {this} constraint does not materially affect the results, as the model remains flexible enough to recover all relevant states.

\section{Simulation study}
\label{sec:simstud}

To evaluate the impact of the different initialization strategies on the performance of the beam sampler in the multivariate Gaussian iHMM, we rely on two different simulation experiments. 
The first investigates the performance under model correctness, where the data-generating process follows a multivariate Gaussian HMM. The second examines the robustness to model misspecification, using data generated from a multivariate Student-$t$ HMM with heavy tails, specifically with $\nu = 5$ degrees of freedom. 
These two experiments allow us to assess the robustness of both the model and the initialization procedures.

We generate synthetic datasets from a finite-state HMM with $K$ latent states. Each regime is characterized by its own state-specific mean or location vector $\bm{\mu}_k$ and covariance matrix $\bm{\Sigma}_k$, and the corresponding transition matrix is constructed to control regime persistence. 
Specifically, we fix the diagonal elements of the transition matrix equal to 
\begin{align*}
    \pi_{ii} = 0.95, \qquad i=1,\ldots,K,
\end{align*}
so that each regime is highly persistent, while the off-diagonal elements satisfy
\begin{align*}
    \pi_{ij} = \frac{1-\pi_{ii}}{K-1} = \frac{0.05}{K-1}, \qquad i\ne j.
\end{align*}
This specification ensures a realistic level of regime persistence without inducing degenerate transitions. 
Emission distributions are drawn from a multivariate Gaussian distribution (for the correctly specified scenario, see Section~\ref{subsec:Gauss}) or from a multivariate Student-$t$ distribution with $\nu = 5$ degrees of freedom (for the misspecified scenario, see Section~\ref{subsec:Studt}), which represent a heavy-tailed behavior as often observed in the economic or financial applications.

\begin{figure}[ht]
    \centering
    \includegraphics[width=0.8\linewidth]
    {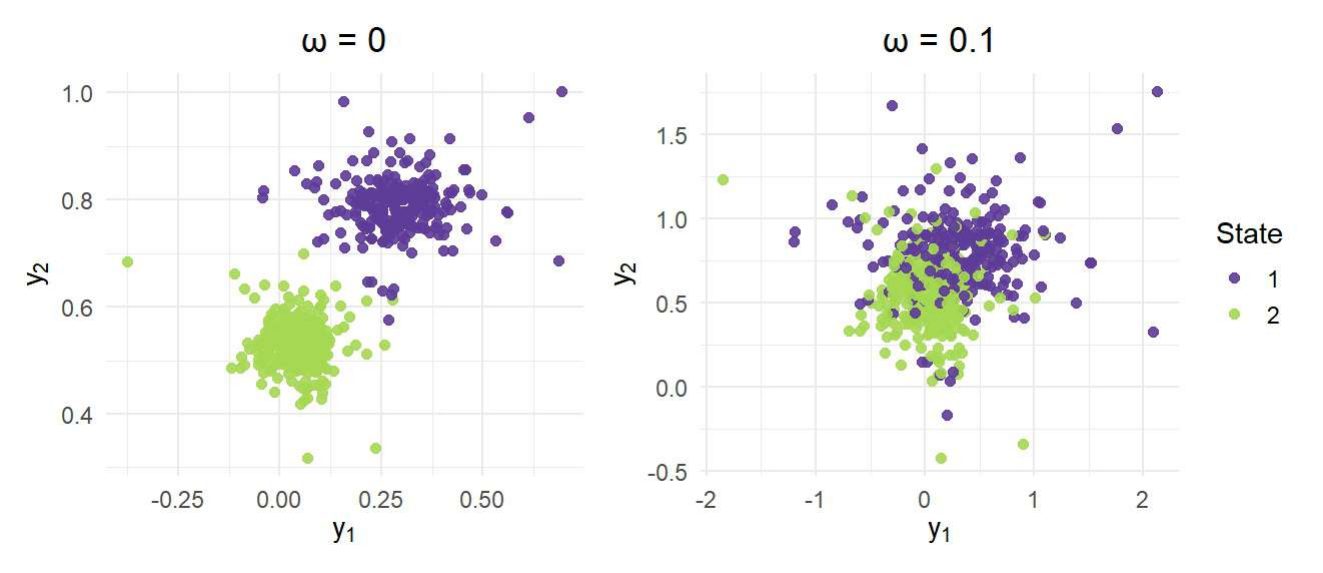}
    \caption{
        Effect of the overlap parameter $\omega$ on the separation between clusters for data simulated from a Student-$t$ distribution with $\nu = 5$, $K = 2$, $T = 500$, and $P = 2$.  
    }
    \label{fig:simdatoverlap}
\end{figure}

To explore a wide range of data-generating conditions, we vary 
\begin{align*}
    \omega \in \{0,0.10\}, \quad K \in \{2,4\}, \quad T \in \{500,1\,000\}, \quad P \in \{5,20\},
\end{align*}
where $K$ represents the number of states, $P$ the number of variables and $T$ the time dimensionality of the dataset. 
Moreover, $\omega$ is the overlap measure, which quantifies the degree of separation between regimes, 
{essentially controlling for the expected overlap probability between distinct clusters \citep{maitra2010simulating}.
}
{
As an example, consider the case of two Gaussian components,
and define 
the pairwise overlap as
$
\omega_{ij} = \omega_{j|i} + \omega_{i|j},
$
where
\[
\omega_{j|i} 
= \mathbb{P}_{X\sim \mathcal N(\boldsymbol{\mu}_i,\boldsymbol{\Sigma}_i)}
\!\left(
\pi_i \phi(X;\boldsymbol{\mu}_i,\boldsymbol{\Sigma}_i)
<
\pi_j \phi(X;\boldsymbol{\mu}_j,\boldsymbol{\Sigma}_j)
\right),
\]
\[
\omega_{i|j} 
= \mathbb{P}_{X\sim \mathcal N(\boldsymbol{\mu}_j,\boldsymbol{\Sigma}_j)}
\!\left(
\pi_j \phi(X;\boldsymbol{\mu}_j,\boldsymbol{\Sigma}_j)
<
\pi_i \phi(X;\boldsymbol{\mu}_i,\boldsymbol{\Sigma}_i)
\right).
\]
The \textit{average} overlap is then computed as
$
\omega 
= \frac{2}{K(K-1)}
\sum_{i<j} \omega_{ij}.
$
We make use of the \textbf{MixSim} \texttt{R} package \citep{melnykov2012mixsim}, 
which generates mean vectors \(\boldsymbol{\mu}_k\) and covariance matrices 
\(\boldsymbol{\Sigma}_k\) matching the desired overlap level \(\omega\) via 
random initialization followed by iterative rescaling of the covariance 
matrices until the target overlap is achieved.
This design avoids reliance on a single covariance configuration and introduces controlled variability in cluster shape and dispersion, yielding a more robust assessment of performance across different covariance structures.
}
Figure~\ref{fig:simdatoverlap} provides an illustration of how $\omega$ influences cluster uncertainty, highlighting that a smaller $\omega$ yields well-separated clusters (left panel), whereas a larger value indicates greater overlap (right panel).

For each combination of $\omega$, $K$, $T$ and $P$, we generate $50$ independent replications, resulting in a total of $800$ datasets for each initialization method. 
Each dataset is then fitted with an iHMM using the four initialization strategies described in Section~\ref{sec:init} (uniform, $k$-means, pam, and mixtures). Unless otherwise stated, we employ $1\,500$ iterations of the beam sampler without burn-in or thinning to assess the convergence speed. 

We evaluate the classification accuracy and the convergence speed
of each initialization method 
by employing the Adjusted Rand Index \citep[ARI,][]{hubert:1985} computed between true and estimated sequences of latent states.\footnote{{See the Supplementary Material for a detailed definition of the ARI.}} 
In our simulation study, 
{for each configuration of the simulation parameters $T, P, K$ and $\omega$,}
we summarize the ARI across seeds,  iterations, {and across all remaining simulation settings not fixed in the given configuration,}
and report the summary statistics at convergence of the beam sampler.

Additionally, in Section~\ref{subsec:convanalysis}, we examine the convergence of the beam sampler under each initialization method using the \citet{Geweke1992} convergence diagnostic and the autocorrelation time (ACT), which assess the similarity of early and late samples and the level of serial dependence in the Markov Chain Monte Carlo (MCMC) draws. 

\subsection{First scenario: Gaussian data}
\label{subsec:Gauss}

The first scenario examines the performance of the four  initialization strategies when the data are generated from a HMM with Gaussian emission distributions. This setting corresponds to a correctly specified model and allows us to evaluate convergence dynamics and classification accuracy under ideal conditions.

\begin{figure}[h!]
    \centering
    \includegraphics[width=.8\linewidth]{ 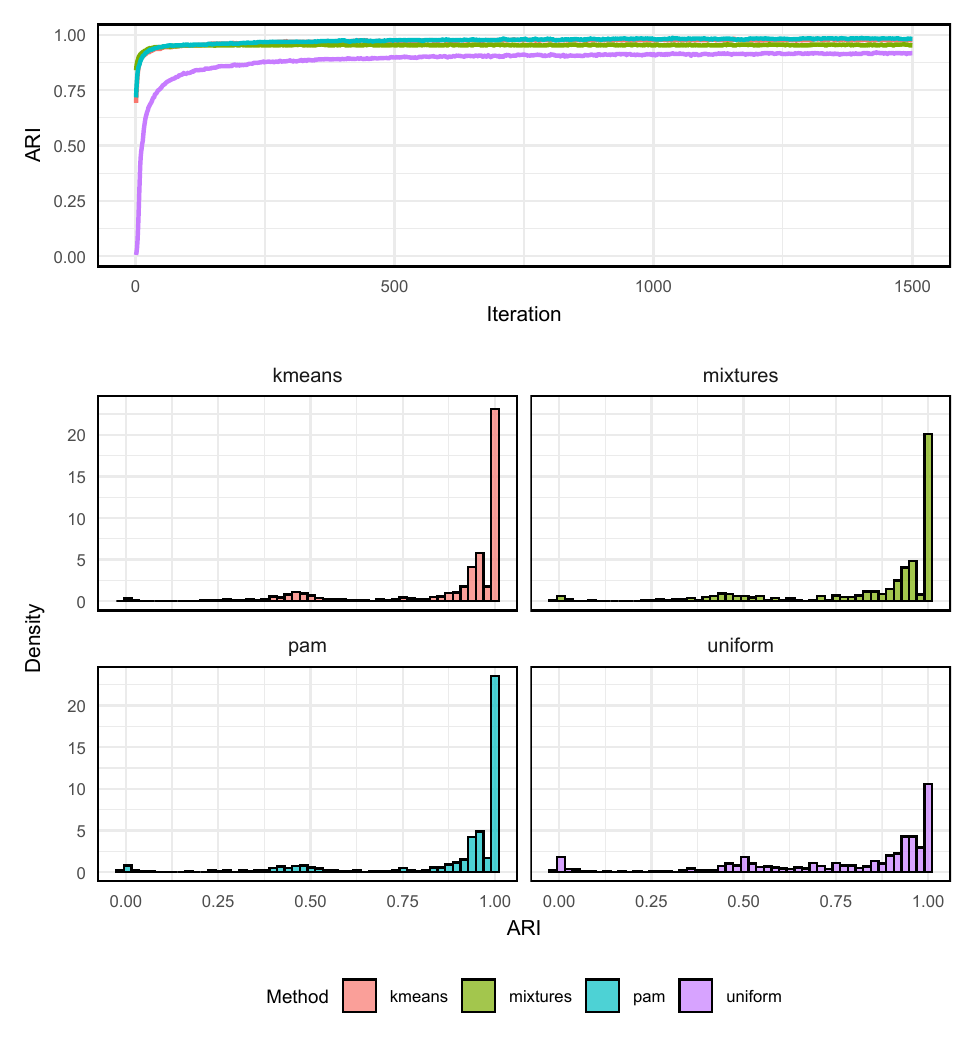}
 \caption{Median ARI across 50 seeds at each iteration computed between true and estimated latent sequences (top panel) and histograms of ARI values at convergence (bottom) for all initialization methods. Results refer to data simulated from a Gaussian distribution. }
    \label{fig:ari_summary}
\end{figure}

Figure~\ref{fig:ari_summary} reports the evolution of the median ARI across $1\,500$ 
iterations (top panel) and the distribution of ARI values at convergence (bottom panel)\footnote{We provide additional results in the Supplementary Material.}.
All initialization strategies show improvements in ARI over iterations but their rates of convergence differ substantially.
The $k$-means, pam, and Gaussian mixture initializations achieve rapid increases in ARI during the first iterations, reaching perfect (ARI around 1) classification in most replications. 
However, the uniform initialization converges slowly and shows greater variability across simulation replications, reflecting the instability of purely random starting partitions. 

Table~\ref{tab:ari_omega_K} compares results for $\omega = 0$ and $\omega = 0.10$, corresponding to clearly and weakly separated clusters, respectively.
In low-overlap conditions ($\omega = 0$), all methods recover the correct latent sequence, yielding ARI values close to $1$, with
$k$-means and pam approaches displaying lower variance across replications.
However, as overlap increases ($\omega = 0.1$), performance differences become more pronounced. The $k$-means initialization remains the most robust to reduced separation, with median ARI values exceeding $0.9$ and exhibiting narrower confidence intervals and lower ARI variability. 
The mixture-based and the pam initializations deteriorate moderately, while uniform initialization often underestimates the number of correct states and exhibits wide dispersion in ARI values.

We also analyze the impact of the true number of regimes ($K$) by focusing on two and four states.
When $K=2$, all initialization methods exhibit accurate recovery of the true number of states, but uniform initialization tends to overestimate the number of states creating extra regimes. 
When $K=4$, the complexity of the latent structure increases, and the uniform and mixture initializations show higher dispersion in the estimated number of regimes, whereas $k$-means and pam continue to recover the correct $K$ with high consistency.  
This suggests that more structured and non-parametric initialization methods might help the sampler to avoid local posterior modes associated with redundant states.

\begin{table}[h!]
\centering
\footnotesize
\caption{
Median, 2.5\% and 97.5\% quantiles and standard deviation (SD) of ARI, and estimated number of states $\hat{K}$, across initialization methods,
for different $\omega$ and $K$, with data simulated from a Gaussian distribution.
Highest ARI median and lowest ARI standard deviation are highlighted in bold.
}
\label{tab:ari_omega_K}

\begin{tabular}{@{}lcccc@{\hskip 1.2em}ccccc@{}}
\toprule
Method & Median & 2.5\% & 97.5\% & SD & Median & 2.5\% & 97.5\% & SD & $\hat{K}$ \\
\midrule
\multicolumn{5}{c}{\textbf{$\omega = 0$}} & \multicolumn{5}{c}{\textbf{$K = 2$}} \\
\cmidrule(lr){2-5} \cmidrule(lr){6-10} 
$k$-means & \textbf{1.00} & 0.97 & 1.00 & 0.01 & 0.98 & 0.90 & 1.00 & 0.06 & \textbf{2.0} \\
mixtures  & \textbf{1.00} & 0.66 & 1.00 & 0.09 & \textbf{1.00} & 0.90 & 1.00 & \textbf{0.04} & \textbf{2.0} \\
pam       & \textbf{1.00} & 1.00 & 1.00 & \textbf{0.00} & 0.98 & 0.87 & 1.00 & 0.13 & \textbf{2.0} \\
uniform   & 0.98 & 0.48 & 1.00 & 0.21 & 0.96 & 0.00 & 1.00 & 0.25 & 3.0 \\
\cmidrule(lr){2-5} \cmidrule(lr){6-10} 
\multicolumn{5}{c}{\textbf{$\omega = 0.10$}} & \multicolumn{5}{c}{\textbf{$K = 4$}} \\
\cmidrule(lr){2-5} \cmidrule(lr){6-10} 
$k$-means & \textbf{0.92} & 0.33 & 0.98 & \textbf{0.23} & \textbf{0.98} & 0.34 & 1.00 & \textbf{0.24} & \textbf{4.0} \\
mixtures  & 0.91 & 0.00 & 0.98 & 0.26 & 0.85 & 0.00 & 1.00 & 0.27 & 3.0 \\
pam       & 0.91 & 0.00 & 0.98 & 0.29 & \textbf{0.98} & 0.00 & 1.00 & 0.29 & \textbf{4.0} \\
uniform   & 0.88 & 0.00 & 0.98 & 0.33 & 0.69 & 0.00 & 1.00 & 0.28 & 3.5 \\
\bottomrule
\end{tabular}
\end{table}

\begin{figure}[h!]
    \centering
    \includegraphics[width=.8\linewidth]{ 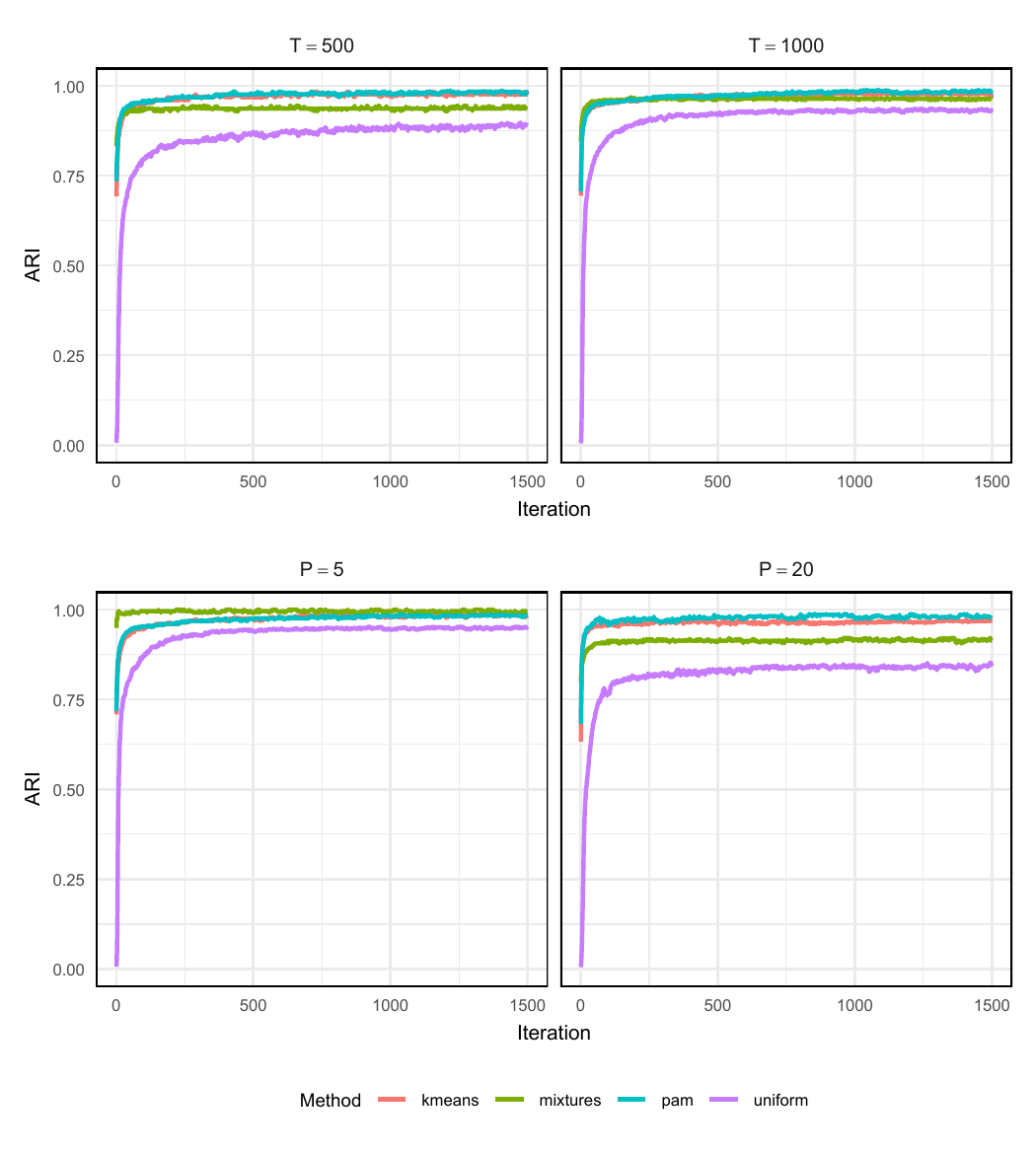}
    \caption{Median ARI across 50 seeds at each iteration, computed between true and estimated latent sequences for all initialization methods for varying $T$ (top panel) and $P$ (bottom). Results refer to data simulated from a Gaussian distribution.}
    \label{fig:varTP}
\end{figure}

Figure~\ref{fig:varTP} displays the median ARI across iterations by focusing on the sequence length $T$ (top panel) and the number of variables $P$ (bottom panel).
When $T$ increases from $500$ to $1\,000$ observations, all initialization schemes benefit from the additional information, achieving smoother ARI trajectories and reducing variability. We notice that uniform initialization always underperforms with respect to the other methods, while the remaining methods seem to reach convergence in the first iterations. 

The results are different when looking at the number of variables $P$. When it increases from $5$ to $20$, both $k$-means and pam maintain high accuracy, whereas mixture initialization decreases, due to the difficulty of estimating full covariance matrices in high-dimensional spaces. 
Uniform initialization deteriorates with the dimensionality, as expected.

In general, we observe that $k$-means achieves higher median ARI values, lower ARI variability, and narrower 95\% confidence intervals across all scenarios, including the most challenging ones, such as $\omega = 0.1$ or a high number of observed variables. On the other hand, the uniform initialization, while simple and widely used as a default method, consistently yields the slowest convergence and highest variability across replications.

\subsection{Second scenario: Student-$t$ data}
\label{subsec:Studt}

\begin{figure}[h!]
    \centering
    \includegraphics[width=.8\linewidth]{ 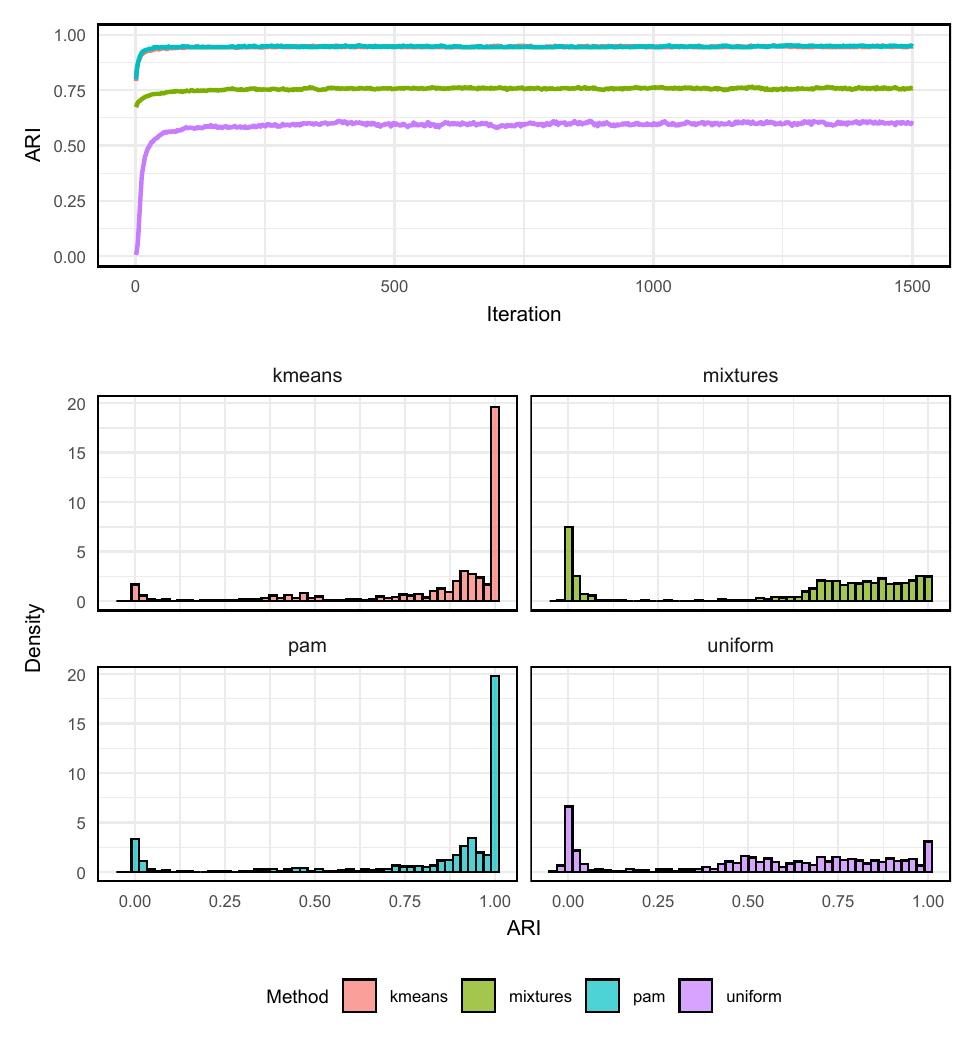}
    \caption{Median ARI across 50 seeds at each iteration computed between true and estimated latent sequences (top panel) and histograms of ARI values at convergence (bottom) for all initialization methods. Results refer to data simulated from a Student-$t$ distribution.}
    \label{fig:ari_summary_Studt}
\end{figure}

The second simulation scenario employs data generated from a Student-$t$ HMM with $\nu=5$ degrees of freedom. 
This design introduces heavy-tailed behavior and within-state variability representing a realistic dataset of model misspecification, which usually arises in macroeconomic applications.
Figure~\ref{fig:ari_summary_Studt} illustrates the evolution of the median ARI across iterations (top panel) and the histogram of ARI values at convergence (bottom panel) for different initialization strategies.
The $k$-means and pam initializations converge rapidly and consistently to the correct clustering, demonstrating strong performance with median ARI above $0.9$.
In contrast, the mixtures and uniform initializations often fail to recover the true partition as shown by low median ARI values. {These} results are also highlighted by the distribution of estimated ARIs in the bottom panel, where the ARI histograms exhibit large dispersion with values in the left tail of the distribution.

Table~\ref{tab:ari_studentT_omega_K} reports results for different overlap values ($\omega = \{0,0.1\}$) and true regime configurations ($K=\{2,4\}$).
When the clusters are clearly separated (top-left panel), $k$-means and pam achieve nearly perfect classification, while mixture initialization shows a drop in the median ARI (0.86), as expected, due to its strong connection with the normality assumption. 
As for the Gaussian scenario, the uniform initialization has the lowest median ARI and highest variability.

When increasing the value of $\omega$,  all methods experience reduced performance due to heavier tails, but the magnitude differs substantially.
The $k$-means and pam initializations achieve median ARI values close to 0.8, whereas the remaining methods deteriorate substantially, with the mixture initialization performing the worst (median ARI equal to 0.29).
\begin{table}[h!]
\centering
\footnotesize
\caption{
Median, 2.5\% and 97.5\% quantiles and standard deviation (SD) of ARI, and estimated number of states $\hat{K}$, across initialization methods,
for different $\omega$ and $K$, with data simulated from a Student-$t$ distribution.
Highest ARI median and lowest ARI standard deviation are highlighted in bold.
}
\label{tab:ari_studentT_omega_K}

\begin{tabular}{@{}lcccc@{\hskip 1.2em}ccccc@{}}
\toprule
Method & Median & 2.5\% & 97.5\% & SD & Median & 2.5\% & 97.5\% & SD & $\hat{K}$ \\
\midrule
\multicolumn{5}{c}{\textbf{$\omega = 0$}} & \multicolumn{5}{c}{\textbf{$K = 2$}} \\
\cmidrule(lr){2-5} \cmidrule(lr){6-10} 
$k$-means & \textbf{1.00} & 0.73 & 1.00 & \textbf{0.07} & \textbf{0.95} & 0.57 & 1.00 & \textbf{0.16} & \textbf{2} \\
mixtures  & 0.86 & 0.56 & 1.00 & 0.13 & 0.72 & 0.00 & 0.96 & 0.34 & 3 \\
pam       & \textbf{1.00} & 0.73 & 1.00 & \textbf{0.07} & \textbf{0.95} & 0.00 & 1.00 & 0.22 & \textbf{2} \\
uniform   & 0.69 & 0.00 & 1.00 & 0.29 & 0.67 & 0.00 & 1.00 & 0.36 & 4 \\
\cmidrule(lr){2-5} \cmidrule(lr){6-10} 
\multicolumn{5}{c}{\textbf{$\omega = 0.10$}} & \multicolumn{5}{c}{\textbf{$K = 4$}} \\
\cmidrule(lr){2-5} \cmidrule(lr){6-10} 
$k$-means & \textbf{0.84} & 0.00 & 0.97 & \textbf{0.31} & 0.94 & 0.00 & 1.00 & \textbf{0.33} & \textbf{4} \\
mixtures  & 0.29 & 0.00 & 0.92 & 0.39 & 0.85 & 0.00 & 1.00 & 0.38 & \textbf{4} \\
pam       & 0.82 & 0.00 & 0.97 & 0.36 & \textbf{0.95} & 0.00 & 1.00 & 0.36 & \textbf{4} \\
uniform   & 0.48 & 0.00 & 0.95 & 0.35 & 0.52 & 0.00 & 1.00 & \textbf{0.33} & \textbf{4} \\
\bottomrule
\end{tabular}
\end{table}

\begin{figure}[h!]
    \centering
    \includegraphics[width=.8\linewidth]{ 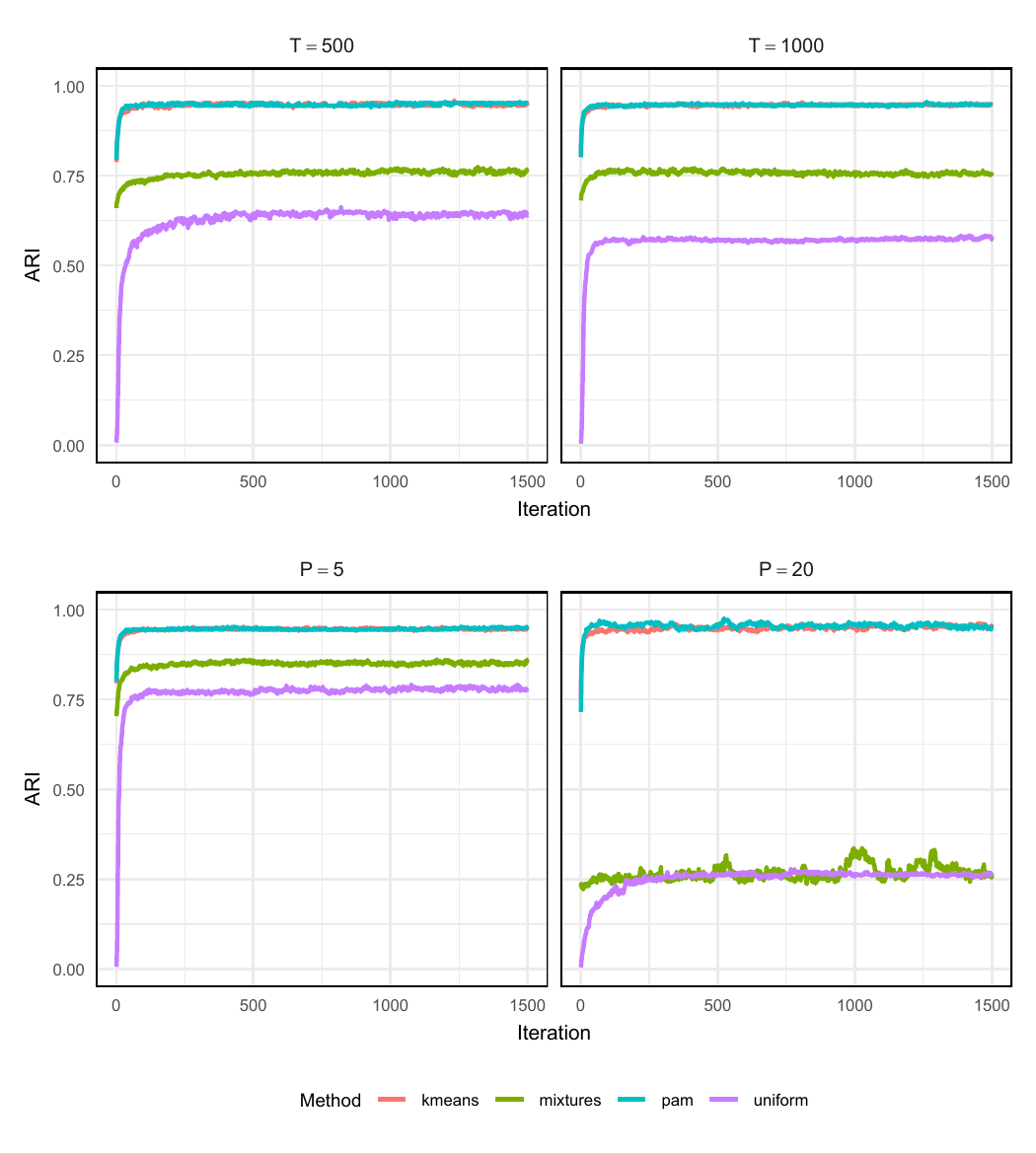}
    \caption{Median ARI across 50 seeds at each iteration, computed between true and estimated latent sequences for all initialization methods for varying $T$ (top panel) and $P$ (bottom). Results refer to data simulated from a Student-$t$ distribution.}
    \label{fig:varTP_Studt}
\end{figure}

We also examine the effect of the true number of latent states $K$ on initialization performance.
When $K = 2$, both mixture and uniform initializations overestimate the number of states. In contrast, $k$-means and pam correctly estimate $\hat{K}$ across replications.
When we increase the number of states to $4$, we notice that all the schemes correctly estimate the regimes, but the median ARI differs among methods, highlighting the good performance of $k$-means and pam with respect to mixtures and uniform. 
Figure~\ref{fig:varTP_Studt} shows the median ARI across iterations when varying the sequence length $T$ (top panel) and the number of variables $P$ (bottom panel).
As $T$ increases from $500$ to $1\,000$,
$k$-means and pam remain consistent and reach convergence, while mixture and uniform initializations become less reliable with higher dimensionality. 
Interestingly, increasing the number of variables $P$ (from $5$ to $20$) slightly favors $k$-means and pam in terms of accuracy. 

To sum up, when the emission distribution deviates from a Gaussian one, uniform or mixture initializations fail to provide reliable starting partitions. 
As for the Gaussian scenario, $k$-means and pam consistently recover the highest median ARI and are similar across  different overlaps, variable and time dimensionality. 

\subsection{Convergence analysis}
\label{subsec:convanalysis}

To complement the classification evaluation of the initialization schemes, we perform different analyses of chain convergence for the beam sampler across all the simulation settings.
We assess convergence by using the classical convergence diagnostic tools and we focus on tests within a chain and across different chains.

Each iHMM model is estimated for $1\,500$ iterations, without thinning, with the first $500$ discarded as burn-in.
For each combination of seed, $T$, $P$, $K$, and $\omega$, we employ the \cite{Geweke1992} diagnostic: for every parameter of the iHMM, we record the median success rate, defined as the share of parameters for which the absolute value of the test-statistic is lower than 2. 
We also compute 50\% and 97.5\% quantiles of autocorrelation time (ACT) across all sampled parameters, with values close to one indicating good mixing and independence of draws. 
Both metrics are then averaged over all runs for each initialization method and configuration.

Table~\ref{tab:geweke_act_omega_K} reports the mean and standard deviation of the Geweke success rate, together with the median and 97.5\% quantile of the ACT, for different $\omega$ and $K$ values under Gaussian emissions. 
Overall, all initialization methods exhibit high Geweke success rates, indicating satisfactory convergence. 
The upper quantile of the ACT is considerably higher for the uniform initialization—particularly when $\omega = 0.10$ and $K = 2$, while it remains below 2 for the other methods across all settings.  

The advantage of $k$-means and pam can be attributed to their ability to generate structured and informative initial partitions that closely approximate the posterior mode. 
These initializations provide the beam sampler with a well-defined starting point, reducing the need for early-stage reallocation of latent states.

\begin{table}[h!]
\centering
\footnotesize
\caption{
Mean Geweke success rate and median autocorrelation time (ACT) across initialization methods, 
with standard deviation of Geweke and 97.5\% quantile of ACT reported below each estimate, respectively, 
for different $\omega$ and $K$ values under Gaussian emissions.}
\label{tab:geweke_act_omega_K}

\begin{tabular}{@{}lcccccccc@{}}
\toprule
Method 
& Geweke & ACT 
& Geweke & ACT 
& Geweke & ACT 
& Geweke & ACT \\
\midrule
& \multicolumn{2}{c}{$\omega = 0$} 
& \multicolumn{2}{c}{$\omega = 0.10$} 
& \multicolumn{2}{c}{$K = 2$} 
& \multicolumn{2}{c}{$K = 4$} \\
\cmidrule(lr){2-3} \cmidrule(lr){4-5} \cmidrule(lr){6-7} \cmidrule(lr){8-9}

$k$-means 
& \makecell{0.93 \\ \scriptsize (0.08)} 
& \makecell{1.00 \\ \scriptsize (1.00)}
& \makecell{0.83 \\ \scriptsize (0.12)} 
& \makecell{1.00 \\ \scriptsize (1.81)}
& \makecell{0.88 \\ \scriptsize (0.11)} 
& \makecell{1.00 \\ \scriptsize (1.59)}
& \makecell{0.88 \\ \scriptsize (0.12)} 
& \makecell{1.00 \\ \scriptsize (1.33)} \\

mixtures  
& \makecell{0.94 \\ \scriptsize (0.04)} 
& \makecell{1.00 \\ \scriptsize (1.00)}
& \makecell{0.84 \\ \scriptsize (0.11)} 
& \makecell{1.00 \\ \scriptsize (1.32)}
& \makecell{0.90 \\ \scriptsize (0.08)} 
& \makecell{1.00 \\ \scriptsize (1.14)}
& \makecell{0.88 \\ \scriptsize (0.11)} 
& \makecell{1.00 \\ \scriptsize (1.31)} \\

pam       
& \makecell{0.94 \\ \scriptsize (0.06)} 
& \makecell{1.00 \\ \scriptsize (1.00)}
& \makecell{0.81 \\ \scriptsize (0.15)} 
& \makecell{1.00 \\ \scriptsize (1.88)}
& \makecell{0.88 \\ \scriptsize (0.12)} 
& \makecell{1.00 \\ \scriptsize (1.37)}
& \makecell{0.87 \\ \scriptsize (0.13)} 
& \makecell{1.00 \\ \scriptsize (1.35)} \\

uniform   
& \makecell{0.85 \\ \scriptsize (0.16)} 
& \makecell{1.00 \\ \scriptsize (1.21)}
& \makecell{0.77 \\ \scriptsize (0.17)} 
& \makecell{1.00 \\ \scriptsize (6.01)}
& \makecell{0.79 \\ \scriptsize (0.18)} 
& \makecell{1.00 \\ \scriptsize (5.24)}
& \makecell{0.83 \\ \scriptsize (0.16)} 
& \makecell{1.00 \\ \scriptsize (1.68)} \\

\bottomrule
\end{tabular}
\end{table}

Table~\ref{tab:geweke_act_omega_K_studt} presents the corresponding convergence diagnostics for data generated from a Student-$t$ distribution. 
The $k$-means and pam initializations maintain high Geweke success rates when $\omega = 0$, though these decrease as cluster overlap increases. 
Median ACT values remain close to one for all methods, but the 97.5\% quantiles are notably higher across the board: this fact highlights the importance of running multiple chains to ensure the reliability and convergence of the results, especially when the Gaussianity assumption of the data-generating process is not guaranteed.

\begin{table}[h!]
\centering
\footnotesize
\caption{
Mean Geweke success rate and median autocorrelation time (ACT) across initialization methods, 
with standard deviation of Geweke and 97.5\% quantile of ACT reported below each estimate, respectively, 
for different $\omega$ and $K$ values under Student-$t$ emissions.}
\label{tab:geweke_act_omega_K_studt}

\begin{tabular}{@{}lcccccccc@{}}
\toprule
Method 
& Geweke & ACT 
& Geweke & ACT 
& Geweke & ACT 
& Geweke & ACT \\
\midrule
& \multicolumn{2}{c}{$\omega = 0$} 
& \multicolumn{2}{c}{$\omega = 0.10$} 
& \multicolumn{2}{c}{$K = 2$} 
& \multicolumn{2}{c}{$K = 4$} \\
\cmidrule(lr){2-3} \cmidrule(lr){4-5} \cmidrule(lr){6-7} \cmidrule(lr){8-9}

$k$-means 
& \makecell{0.92 \\ \scriptsize (0.10)} 
& \makecell{1.00 \\ \scriptsize (1.84)}
& \makecell{0.73 \\ \scriptsize (0.20)} 
& \makecell{1.11 \\ \scriptsize (7.03)}
& \makecell{0.83 \\ \scriptsize (0.18)} 
& \makecell{1.00 \\ \scriptsize (2.75)}
& \makecell{0.82 \\ \scriptsize (0.18)} 
& \makecell{1.00 \\ \scriptsize (5.76)} \\

mixtures  
& \makecell{0.78 \\ \scriptsize (0.18)} 
& \makecell{1.13 \\ \scriptsize (3.80)}
& \makecell{0.82 \\ \scriptsize (0.16)} 
& \makecell{1.22 \\ \scriptsize (4.30)}
& \makecell{0.81 \\ \scriptsize (0.18)} 
& \makecell{1.33 \\ \scriptsize (4.46)}
& \makecell{0.80 \\ \scriptsize (0.17)} 
& \makecell{1.11 \\ \scriptsize (3.99)} \\

pam       
& \makecell{0.91 \\ \scriptsize (0.10)} 
& \makecell{1.00 \\ \scriptsize (2.08)}
& \makecell{0.74 \\ \scriptsize (0.21)} 
& \makecell{1.12 \\ \scriptsize (7.60)}
& \makecell{0.83 \\ \scriptsize (0.18)} 
& \makecell{1.00 \\ \scriptsize (3.44)}
& \makecell{0.83 \\ \scriptsize (0.19)} 
& \makecell{1.00 \\ \scriptsize (3.31)} \\

uniform   
& \makecell{0.81 \\ \scriptsize (0.18)} 
& \makecell{1.00 \\ \scriptsize (3.08)}
& \makecell{0.78 \\ \scriptsize (0.17)} 
& \makecell{1.12 \\ \scriptsize (6.57)}
& \makecell{0.79 \\ \scriptsize (0.18)} 
& \makecell{1.11 \\ \scriptsize (4.99)}
& \makecell{0.80 \\ \scriptsize (0.17)} 
& \makecell{1.00 \\ \scriptsize (3.49)} \\

\bottomrule
\end{tabular}
\end{table}

In conclusion, $k$-means and pam show higher Geweke success rates and lower ACT across both Gaussian and non-Gaussian data, while uniform initialization provides slow convergence and appears to be the most problematic scheme.

\section{Applications}
\label{sec:empstud}

In this section, we illustrate the empirical performance of the iHMM on two real-data datasets. 
The first application refers to the monthly industrial production volume index, a standard macroeconomic indicator of business-cycle fluctuations, for nine European countries. This dataset is useful for exploring regime switches since they may refer to business-cycle phases or major economic shocks.
{
The second application analyses daily log-returns of six financial assets referring to different market sectors. This dataset provides a more complex scenario in which to test iHMM performance based on the different initialization strategies.}

For each dataset, we estimate the iHMM using 10 independent initializations  per method ($k$-means, pam, Gaussian mixtures, uniform), and we set vague priors for the hyperparameters $\alpha$ and $\gamma$ as described in Section \ref{sec:iHMM}.
Each chain is run for $5\,000$ iterations, discarding the first $3\,000$ as burn-in. 
Posterior samples from the remaining draws are used to compute parameter summaries and regime probabilities.

We first perform convergence diagnostics at the chain level, following the same procedure described in Section~\ref{subsec:convanalysis}. We consider a chain convergent when it provides a Geweke success rate above 0.75 and a median ACT below 2. 
We then assess convergence across chains using the potential scale reduction factor $\hat{R}$~\citep{GelmanRubin1992}, where values of $\hat{R}$ close to 1 indicate good mixing and cross-chain convergence.

For the converged chains, we pool posterior draws to infer parameter estimates and regime probabilities, while obtaining temporal clustering through the \textit{maximum a posteriori} (MAP) state assignments. 

\subsection{Industrial Production index data}
\label{subsec:prodind}

We investigate the monthly industrial production volume index for nine European economies -- Belgium, Germany, Spain, France, Italy, Lithuania, Hungary, Austria, and Romania -- spanning from\footnote{In the Supplementary Material, we run the same analysis by excluding the COVID-19 outbreak.} March 2002 to August 2025 ($T=282$). 
Each series, drawn from the Eurostat database\footnote{The data are available at 
\href{https://ec.europa.eu/eurostat/databrowser/view/sts_inpr_m/default/table?lang=en}
{\url{https://ec.europa.eu/eurostat/databrowser/view/sts_inpr_m/default/table?lang=en}}
},
 is seasonally and calendar adjusted, and reflects output dynamics across three main sectors:  mining and quarrying; manufacturing; energy production (which includes electricity, gas, steam, and air conditioning supply).
 
\begin{figure}[h!]
    \centering
    \includegraphics[width=\linewidth]{ 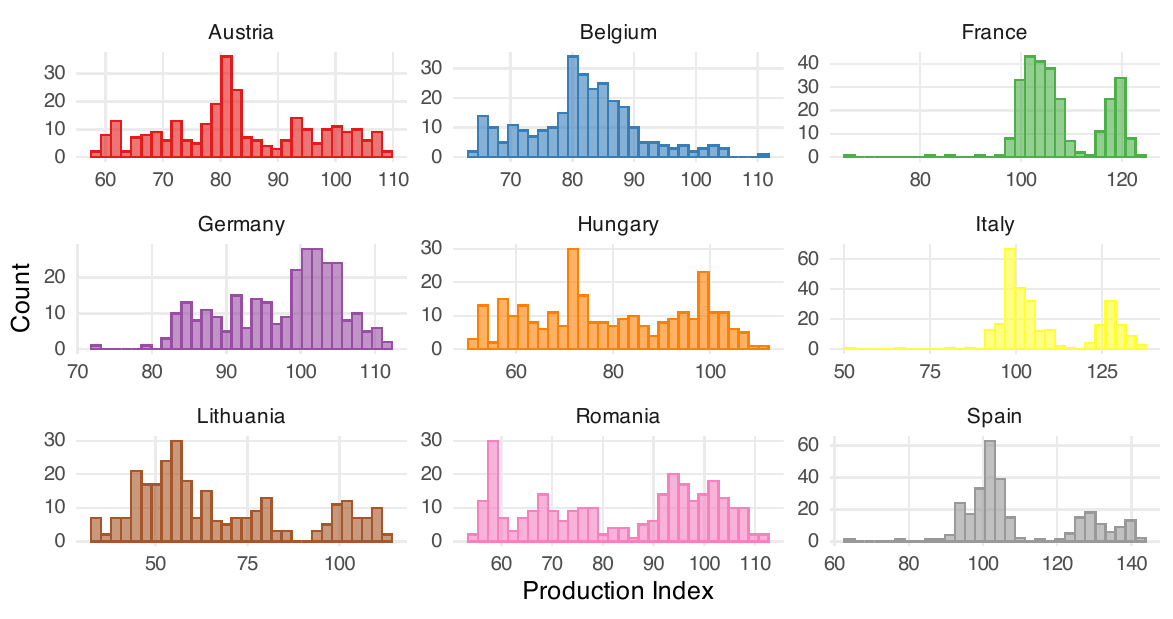}
    \caption{
        Empirical distributions of the industrial production index for nine European countries.
    }
    \label{fig:hists_countries}
\end{figure}

In Figure~\ref{fig:hists_countries}, the cross-country distribution of the monthly series presents clear multimodality and heterogeneous dispersion across countries, signaling potential regime switching.
Such characteristics motivate the use of iHMM to infer the number of unobserved macroeconomic states without restrictive parametric assumptions.

Across initialization methods, the posterior inference consistently favors $\hat{K}=3$, with the exception of the uniform initialization.
{A similar pattern is observed under both $k$-means and pam initializations. In the former, six of ten chains converge with median $\hat{R}$ values close to 1. In the latter, all but one run select three states, and six chains converge with $\hat{R}$ values near 1.}
Close to the results highlighted in the simulation section, the mixture-based initialization also consistently identifies $\hat{K}=3$, but only one chain achieves convergence, suggesting less robust sampling performance for this approach.
In contrast, the uniform initialization predominantly selects $\hat{K}=2$, and none of these chains converge; consequently, and also based on the simulation results, we omit results for this initialization method.

Table \ref{tab:stateCond_var} reports the posterior mean of the traces of the state-conditional covariance matrices. The results indicate consistent regime characterization under both $k$-means and pam initializations, revealing three states with progressively increasing overall volatility. Similarly, Table \ref{tab:stateCond_means} presents the state-conditional mean values for each time-series. Based on the inference obtained using either the $k$-means or pam initialization, states 1 and 2 show broadly similar mean production index values across countries, while state 3 is characterized by  higher heterogeneity, consistent with its association with the high-volatility regime.
\begin{table}[h]
\centering
\caption{Trace of the covariance matrices for each inferred state and initialization method.}
\label{tab:stateCond_var}
\begin{tabular}{lccc}
\toprule
Method & State 1 & State 2 & State 3 \\
\midrule
$k$-means & $233.00$ & $354.00$  & $797.00$ \\
pam       & $215.00$ & $377.00$  & $792.00$ \\
mixtures  & $271.00$ & $2538.00$ & $264.00$ \\
\bottomrule
\end{tabular}
\end{table}

\begin{table}[h]
\centering
\caption{State-conditional mean industrial production indices by initialization method.}
\label{tab:stateCond_means}
\footnotesize
\begin{tabular}{llll lll lll}
\toprule
& \multicolumn{3}{c}{$k$-means}
& \multicolumn{3}{c}{pam}
& \multicolumn{3}{c}{mixtures} \\
\cmidrule(lr){2-4} \cmidrule(lr){5-7} \cmidrule(lr){8-10}
State & 1 & 2 & 3
 & 1 & 2 & 3
 & 1 & 2 & 3 \\
\midrule
Austria   & 101.25 & 86.54 & 71.60 & 101.39 & 86.22 & 71.26 & 87.83 & 83.74 & 77.21 \\
Belgium   & 92.65  & 82.82 & 75.16 & 92.76  & 82.80 & 74.87 & 83.57 & 82.23 & 78.46 \\
France    & 100.39 & 105.27 & 113.11 & 100.52 & 105.26 & 113.37 & 105.60 & 109.93 & 103.69 \\
Germany   & 96.22  & 104.42 & 92.48 & 96.25  & 104.23 & 92.08 & 105.31 & 94.02 & 95.89 \\
Hungary   & 99.74  & 84.29 & 64.34 & 99.87  & 83.67 & 64.09 & 87.45 & 79.52 & 67.84 \\
Italy     & 97.03  & 100.77 & 119.93 & 97.25  & 100.97 & 120.35 & 100.82 & 113.30 & 104.29 \\
Lithuania & 101.34 & 65.96 & 48.72 & 101.54 & 65.35 & 48.60 & 68.74 & 71.33 & 51.17 \\
Romania   & 98.48  & 95.27 & 64.99 & 98.56  & 94.34 & 64.46 & 98.84 & 78.22 & 74.26 \\
Spain     & 101.19 & 98.74 & 122.41 & 101.34 & 98.86 & 123.29 & 99.81 & 117.38 & 101.60 \\
\bottomrule
\end{tabular}
\end{table}

\begin{figure}[!htbp]
    \centering
    \includegraphics[width=.9\linewidth]{ 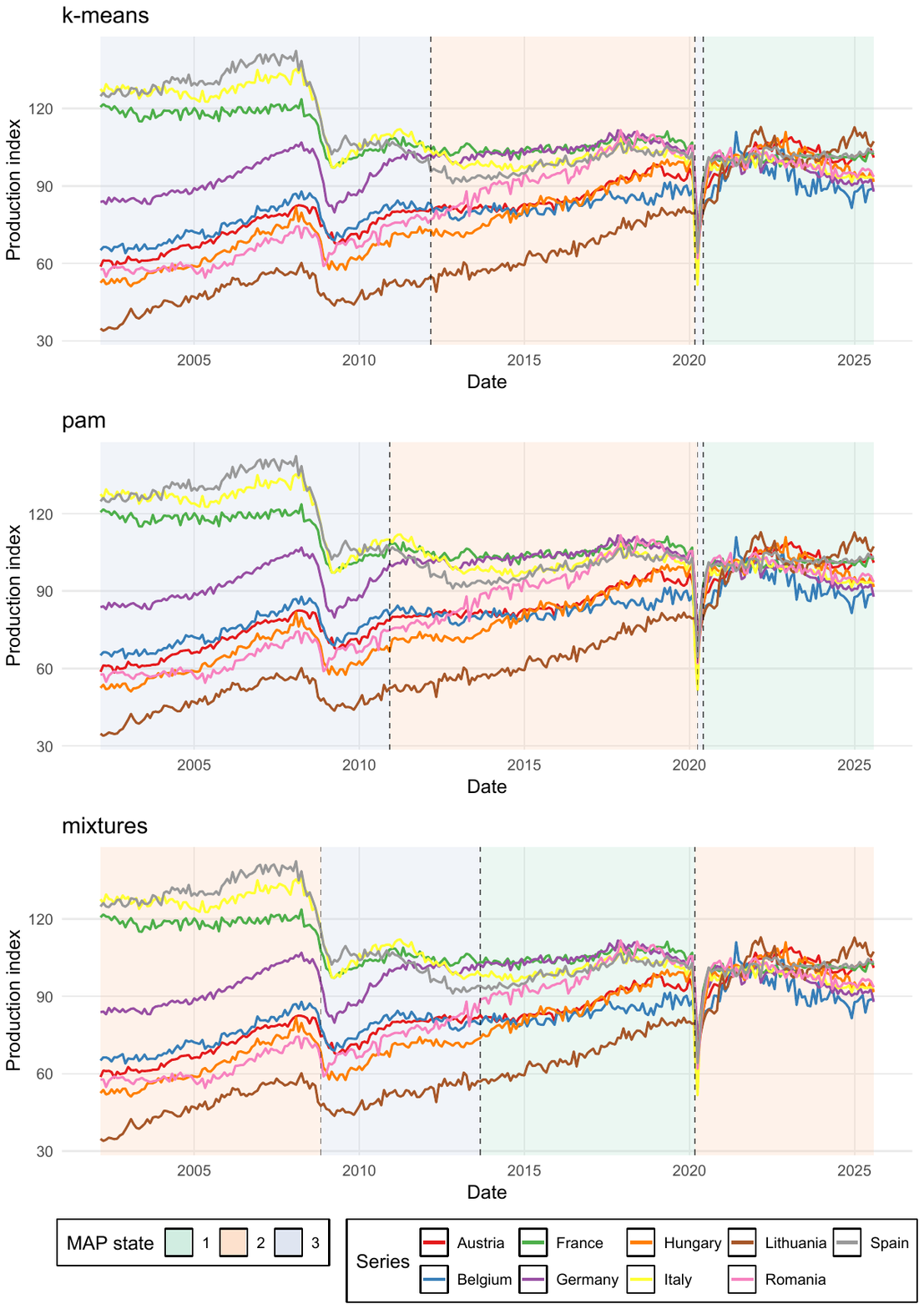}
    \caption{Regime classification over time for each initialization method, and for each country.}
    \label{fig:map_prodind}
\end{figure}

Figure~\ref{fig:map_prodind} illustrates the time-series clustering obtained via MAP classification of the estimated regime probabilities for all initialization methods. Both $k$-means and pam initializations yield highly consistent classifications, as indicated by an ARI of 0.92 computed between the MAP-based state assignment sequences obtained under the two methods.
They both identify four main phases in the European business cycle: an initial high-volatility, heterogeneous phase prior to the Global Financial Crisis (2002--2008), followed by a more stable one (state 2) starting in February 2012 for $k$-means and August 2011 for pam where the economies recovered from the crisis. 
Then a return to high volatility around February 2020, coinciding with the outbreak of COVID-19. 
This turbulent phase lasts two months for $k$-means and one month for pam. The subsequent period is dominated by state 1, characterized by low volatility and similar mean production index levels across countries.

Regarding the inference obtained with the Gaussian mixture initialization, Table \ref{tab:stateCond_var} indicates a different regime structure: state 2 exhibits higher volatility, while the other two display comparable variance levels. 
It is not straightforward to distinguish between states 1 and 3. As shown in Table \ref{tab:stateCond_means}, the largest differences in state-conditional means occur for Lithuania, Hungary, and Romania, which exhibit substantially lower values in state 3. 
Four distinct phases can be identified, as illustrated in Figure \ref{fig:map_prodind}: an initial high-volatility period ending in November 2008, a subsequent moderate-volatility phase (state 3) lasting until August 2013, a stable phase (state 1) persisting up to February 2020, and finally a high-volatility phase until the end of the observation period.

{
\subsection{Financial markets data}
\label{subsec:financial}

Although the industrial production index application yields satisfactory results, the identification of economic phases may appear relatively straightforward even upon visual inspection. 
Thus, we focus on financial time-series, where the appropriate number of latent regimes is typically not discernible by simple visual inspection. 
Figure~\ref{fig:hist_finance} reports the histograms of returns for six financial assets related to different market sectors; indeed, the evidence does not provide a clear indication of how the number of regimes $K$ should be specified.
\begin{figure}[h!]
    \centering
    \includegraphics[width=\linewidth]{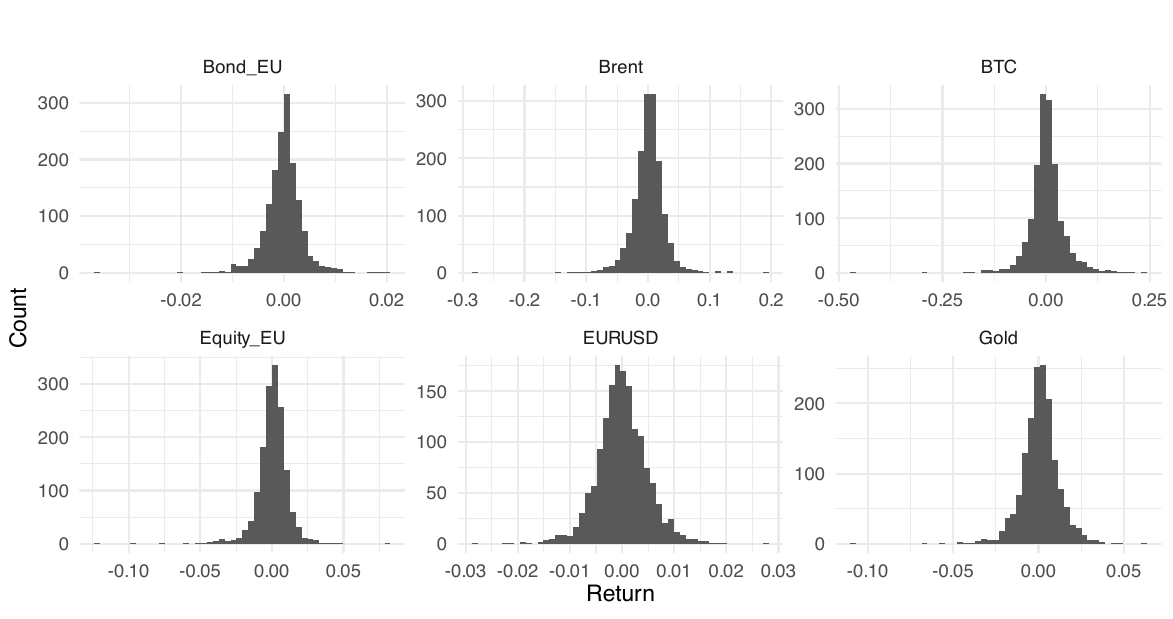}
   \caption{Histograms of daily log-returns for Bond EU (iShares Core Euro Aggregate Bond ETF), Brent crude oil, BTC (Bitcoin in USD), Equity EU (STOXX Europe 600), EURUSD (EUR–USD exchange rate), and Gold.}
    \label{fig:hist_finance}
\end{figure}
We retrieve daily data from Yahoo Finance using the \textbf{quantmod} \texttt{R} package \citep{ryan2020package}, covering the period from January 4, 2019, to March 6, 2026. This time span includes many episodes of financial stress, including the COVID-19 market crash, the 2022 Russian invasion of Ukraine and the associated energy shock, the volatility surrounding U.S. tariff announcements, and recent geopolitical tensions 
in the Middle East. Specifically, we consider the following time-series: the STOXX Europe 600 index (Equity EU), as a proxy for European equity markets; the iShares Core Euro Aggregate Bond ETF (Bond EU), capturing the euro-area fixed-income market; Gold Shares to track gold prices; BTC-USD, the dollar price of Bitcoin; the EUR-USD exchange rate, reflecting major foreign exchange dynamics; and Brent crude oil prices, capturing developments in global energy markets.

Under both $k$-means and pam initializations, all chains converge both in terms of Geweke success rate and ACT, with $\hat{R}\approx1$, and the posterior distributions favor $\hat{K} = 2$. 
Table~\ref{tab:stateCond_returns_kmeans_pam} characterizes the two regimes in terms of state-conditional mean returns:  state~1 is associated with positive returns across all assets, corresponding to a bull regime, whereas state~2 exhibits negative returns, particularly for equity and energy markets, identifying a bear regime. 
\begin{table}[h]
\centering
\caption{State-conditional mean asset returns (\%) under $k$-means and pam initializations.}
\label{tab:stateCond_returns_kmeans_pam}
\footnotesize
\begin{tabular}{lcccc}
\toprule
 & \multicolumn{2}{c}{$k$-means} & \multicolumn{2}{c}{pam} \\
\cmidrule(lr){2-3}\cmidrule(lr){4-5}
Asset & State 1 & State 2 & State 1 & State 2 \\
\midrule
Equity EU & 6.20  & -23.59 & 6.15  & -23.62 \\
Bond EU   & 0.73  & -2.72  & 0.74  & -2.84  \\
Gold       & 3.78  & -14.51 & 3.66  & -14.22 \\
BTC        & 0.69  & -2.67  & 0.68  & -2.44  \\
EURUSD     & 0.55  & -2.08  & 0.44  & -1.77  \\
Brent      & 6.36  & -24.08 & 6.33  & -24.38 \\
\bottomrule
\end{tabular}
\end{table}
The traces of the variance-covariance matrices are identical across methods, and equal to 0.001 and 0.009 in the bull and bear regimes, respectively.

Turning to the transition dynamics, identical under $k$-means and pam initializations, the estimated probability of remaining in the bull state is 0.89, while the probability of switching from bull to bear is 0.11. In contrast, the bear state is less persistent, with self-transition probability equal to 0.57.

Figure~\ref{fig:ts_financial} shows the standardized price series for all six assets alongside the estimated probability of being in the bull regime over time. 
\begin{figure}[h!]
    \centering
    \includegraphics[width=\linewidth]{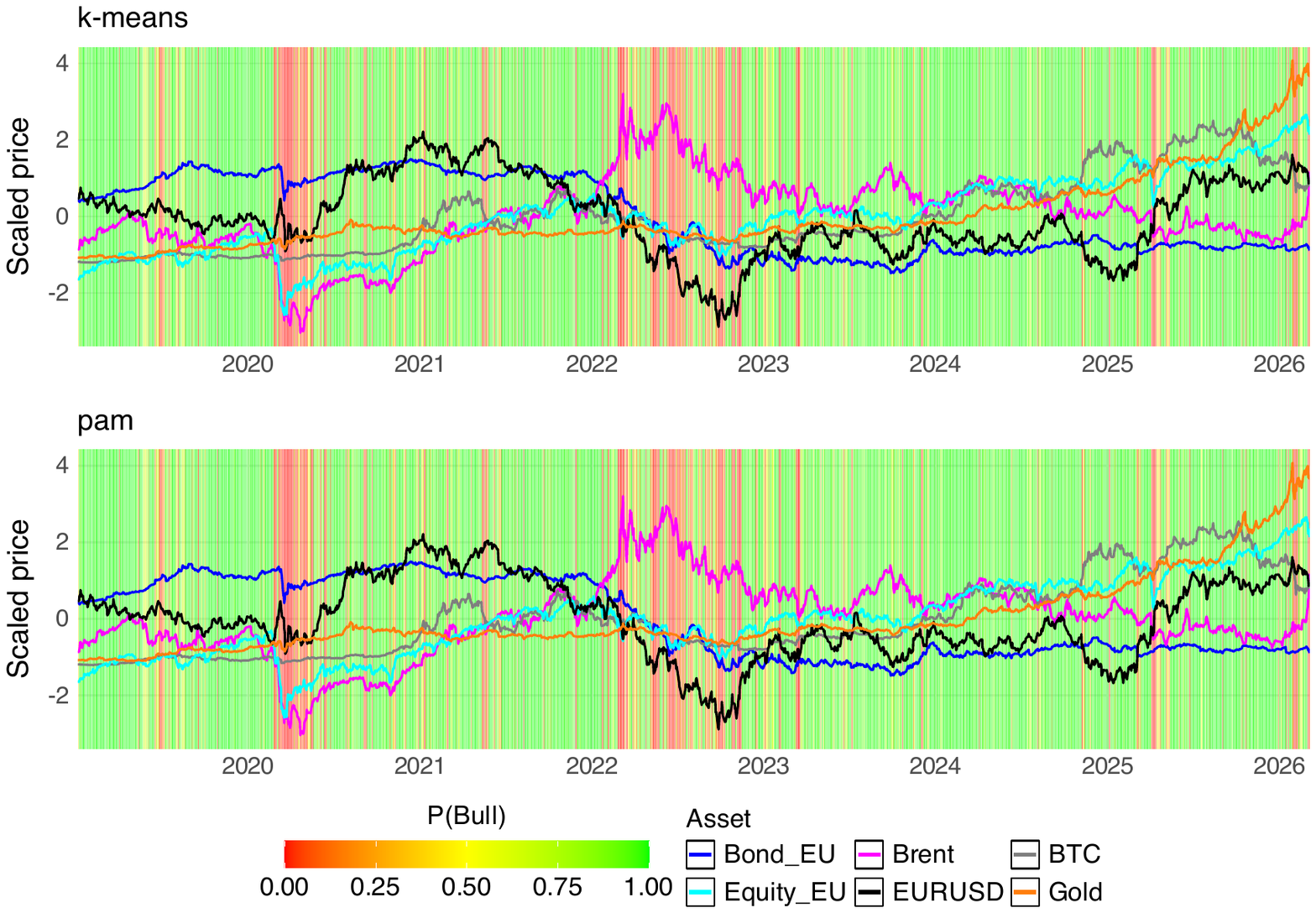}
    \caption{Probability of bull state for each asset, for $k$-means and pam initializations. Scaled prices on the y-axis.}
    \label{fig:ts_financial}
\end{figure}

Under mixture initialization, all chains converge to $\hat{K} = 3$, and only half of them exhibit satisfactory Geweke diagnostics and ACTs. However, the economic interpretation of the resulting regimes is less straightforward: while a clear bear phase emerges, the remaining two states display very similar state-conditional mean returns and volatilities.
Under uniform initialization\footnote{{In the Supplementary Material, we present an application of the iHMM to the Old Faithful geyser dataset, demonstrating that uniform initialization performs poorly even in this simple setting with well-separated clusters.}}, the chains converge to different values of $K$, and none of them achieves satisfactory Geweke statistics or ACT. For these reasons, we do not report results for these two initialization methods.
}

\section{Conclusions}
\label{sec:conclusions}

This study highlights the impact of initialization strategies in the estimation of infinite hidden Markov models using the beam sampler algorithm.
Our results show that the choice of  initialization strategy can substantially influence the quality of the inferred latent structure and the convergence behavior of the model.

Across different simulation experiments, uniform initialization consistently underperforms, failing to recover the correct latent states even when the underlying data are generated from well-separated Gaussian mixtures.
In contrast, initializations based on $k$-means or partitioning around medoids 
yield more accurate and stable estimates.
While Gaussian mixture initialization remains competitive, its performance tends to deteriorate as the feature dimensionality increases or when the assumption of Gaussianity of the data is violated.

The empirical applications reinforce these findings. In the European industrial production dataset, macroeconomic phases are correctly recovered when using $k$-means and pam. Similarly, in the financial application, they both correctly identify bull and bear market phases, and provide sound and interpretable state-conditional parameters.

{
All the experiments indicate that $k$-means and pam achieve high classification accuracy even under model misspecification: this suggests that adopting more flexible emission distributions, such as Student-$t$ or skewed Student-$t$ specifications, may further enhance robustness and state recovery. A systematic investigation of the interaction between initialization strategies and richer emission models represents a natural direction for future research.
}

\section*{Disclosure statement}\label{disclosure-statement}

The authors have no conflicts of interest to declare.

This research did not receive any specific grant from funding agencies in the public, commercial, or not-for-profit sectors.

\section*{Data Availability Statement}\label{data-availability-statement}

The data for the application of Section \ref{subsec:prodind} are available at 
\href{https://ec.europa.eu/eurostat/databrowser/view/sts_inpr_m/default/table?lang=en}
{\url{https://ec.europa.eu/eurostat/databrowser/view/sts_inpr_m/default/table?lang=en}}

The data for the application of Section \ref{subsec:financial} are retrieved from Yahoo Finance using the \textbf{quantmod} package in \texttt{R}.

\bibliography{biblio}
\clearpage

\renewcommand{\thesection}{S\arabic{section}}
\setcounter{section}{0}

{\huge \centering Supplementary Material}

\vspace{3mm}

\noindent
Section \ref{sec:HDP} provides more details on the hierarchical Dirichlet process, while Section \ref{sec:beam} presents a detailed description of the beam sampler for posterior inference.
Section \ref{sec:ARI} shows the definition of the Adjusted Rand Index, and Section \ref{sec:additionalsim} provides additional Tables and Figures from the simulation studies in Section 4 of the main paper. In Section \ref{sec:addemp}, we study the macroeconomic application of the main paper by excluding the COVID-19 outbreak, while in Section \ref{sec:old} we show an application to the well-known Old Faithful geyser dataset.

\section{Hierarchical Dirichlet process}
\label{sec:HDP}

A hierarchical Dirichlet process \citep[HDP,][]{teh2006hierarchical} is defined on a probability space $(\Theta, \mathcal{B})$, considering a set of random probability measures $G_k$, one for each group of observations\footnote{In the iHMM case, one for each row of the transition matrix.}, and a global
random probability measure $G_0$
\begin{align*}
    G_k|\alpha,G_0 &\sim \operatorname{DP}(\alpha,G_0), \\
    G_0 |\gamma, H &\sim \operatorname{DP}(\gamma, H),
\end{align*}
where $H$ is the base measure and $\alpha$ and $\gamma$ are the concentration parameters.

Observations are
organized into groups and assumed to be exchangeable both
\textit{within} each group, $i \in \{1,\ldots, T\}$, and \textit{across} groups, $k \in \{1, \ldots, K\}$.
A HDP can be used as the prior distribution over the mixture component $\theta_{ki}$ associated with
each observation $y_{ki}$ for grouped data. For each $k$, let $\theta_{k1}, \theta_{k2}, \ldots$ be i.i.d. random variables distributed as $G_k$, where each $\theta_{ki}$ corresponds to a single observation $y_{ki}$. 
The likelihood is given by 
$$\theta_{ki}|G_k\sim G_k,$$
$$y_{ki}|\theta_{ki}\sim F(\theta_{ki}).$$
Given that the global measure $G_0$ is distributed as a DP, it can be expressed using the \cite{sethuraman1994constructive} representation
$$G_0=\sum_{k'=1}^{\infty}\beta_{k'} \delta_{\phi_{k'}},$$
with $\phi_{k'}|H\sim H$ independently and $\pmb{\beta}=(\beta_{k'})_{k'=1}^{\infty}\sim \operatorname{GEM}(\gamma)$ (Griffiths, Engen and McCloskey) are mutually independent.
$\pmb{\beta}\sim\operatorname{GEM}(\gamma)$ is obtained through the stick-breaking construction
    $$v_{k'}|\gamma \sim \operatorname{Beta}(1,\gamma),$$
    $$\beta_{k'}=v_{k'}\prod_{l=1}^{k'-1}(1-v_l) \quad k'=1,\ldots,\infty.$$
    The support of each $G_k$ is contained within the support of $G_0$. Thus the stick-breaking representation for $G_j$ is a reweighted sum of the atoms in $G_0$
    $$G_k=\sum_{k'=1}^{\infty}\pi_{kk'}\delta_{\phi_k'}.$$
    Then, we have that
$$\pmb{\pi}_k=(\pi_{k1},\pi_{k2},\ldots)\sim \operatorname{DP}(\alpha,\pmb{\beta}).$$
First note that $\pmb{\pi}_k$ are independent given $\pmb{\beta}$ because the $G_k$'s are independent given $G_0$.
   Let $(A_1,\ldots,A_r)$ be a  measurable partition of $\Theta$ and let $(K_1,\ldots, K_l)$ be a measurable partition of the integers, for $l=1,\ldots,r$, we have 
    $$K_l=\{k':\phi_{k'} \in A_l\}.$$
    Further, assuming that $H$ is nonatomic, the $\phi_{k'}$'s are distinct with probability 1, and so any partition of the positive integers corresponds to some partition of $\Theta$. Thus, for each $k$
    $$
\left(G_{k}\left(A_{1}\right), \ldots, G_{k}\left(A_{r}\right)\right)
\sim \operatorname{Dir}\left(\alpha G_{0}\left(A_{1}\right), \ldots, \alpha G_{0}\left(A_{r}\right)\right)$$
$$\Rightarrow \quad\left(\sum_{k' \in K_{1}} \pi_{k k'}, \ldots, \sum_{k' \in K_{r}} \pi_{k k'}\right)
\sim \operatorname{Dir}\left(\alpha \sum_{k' \in K_{1}} \beta_{k'}, \ldots, \alpha \sum_{k' \in K_{r}} \beta_{k'}\right),$$
for every finite partition of the positive integers. 
Hence, we have that $$\pmb{\pi}_k \overset{\mathrm{iid}}{\sim} \operatorname{DP}(\alpha,\pmb{\beta}).$$
Based on \cite{hjort2010bayesian} and some algebra, this yields the following explicit construction for $\pmb{\pi}_k|\pmb{\beta}$
    $$\begin{aligned}
v_{kk'} \mid \alpha, \beta_{1}, \ldots, \beta_{k'} & \sim \operatorname{Beta}\left(\alpha \beta_{k'}, \alpha\left(1-\sum_{l=1}^{k'} \beta_{l}\right)\right), \\
\pi_{kk'} &=v_{kk'} \prod_{l=1}^{k'-1}\left(1-v_{k' l}\right), \quad \text { for } k'=1, \ldots, \infty .
\end{aligned}$$
\cite{van2008beam} identify each $G_k$ as describing both the transition probabilities $\pi_{kk'}$ from state $k$ to state $k'$ and the emission distribution parametrized by $\phi_{k'}$, so we can formally define the iHMM as
    $$ \pmb{\beta}\sim \operatorname{GEM}(\gamma),$$
    $$\pmb{\pi}_k|\pmb{\beta}\sim\operatorname{DP}(\alpha,\pmb{\beta}) \quad k=1,\ldots, \infty,$$
    $$\phi_k\sim H,$$
    $$s_t|s_{t-1} \sim \operatorname{Multinomial}(\pmb{\pi}_{s_{t-1}}),$$
    $$\textbf{y}_t|s_t\sim F(\phi_{s_t}) \quad t=1,\ldots, T.$$
    Finally, we place priors over the hyperparameters $\alpha$ and $\gamma$, and a common solution uses
    $$\alpha\sim \operatorname{Gamma}(a_\alpha,b_\alpha),$$
    $$\gamma \sim \operatorname{Gamma}(a_\gamma,b_\gamma).$$
    
    \section{The beam sampler}
    \label{sec:beam}
    The terminology reflects its resemblance to \textit{beam search} \citep{lowerre1976harpy}, a heuristic algorithm designed to approximate the maximum a posteriori trajectory, given observations, in nonlinear dynamical systems.
   The underlying idea in both is to limit the search to a small number of states.
    Beam sampling combines \textit{slice sampling} \citep{walker2007sampling} to limit the number of states to a finite set, and \textit{dynamic programming}, specifically the forward-backward algorithm \citep{dempster:1977}, to sample trajectories efficiently.
Beam sampling proceeds by introducing an auxiliary variable so that, conditional on it, only a finite number of trajectories have positive probability.
The forward-backward algorithm can then be used to compute the conditional probabilities of each of these trajectories.
In the first step, initialize the hidden state sequence $$\pmb{s}=(s_1,\ldots,s_T),$$ and the model parameters $\pmb{\pi}$, $\pmb{\beta}, \boldsymbol{\phi}$.
For $t=1,\ldots,T$, we introduce an auxiliary variable $u_t$ with conditional distribution 
    $$u_t|\pmb{\pi},s_{t-1},s_t\sim \mathcal{U}(0,\pi_{s_{t-1}s_t}).$$
    The second step consists of sampling the whole trajectory $\pmb{s}$ given the auxiliary variable $u$ using a form of forward filtering-backward sampling:
    \begin{enumerate}
        \item[2.1] Initialization $$p(s_1=1)=1.$$
        \item[2.2] For $t=2,\ldots,T$
        $$p(s_t|y_{1:t},u_{1:t}) \propto p(y_t|s_t)\sum_{s_{t-1}:u_t<\pi_{s_{t-1}s_t}} p(s_{t-1}|y_{1:t-1},u_{1:t-1}).$$
        \item[2.3] Sample for $t=T$ from
        $$p(s_T|y_{1:T},u_{1:T}).$$
        \item[2.4] For each $t=T-1,\ldots,1$
        $$p(s_t|s_{t+1},y_{1:T},u_{1:T}) \propto p(s_{t+1}|s_t,u_{t+1})p(s_t|y_{1:t},u_{1:t}).$$
    \end{enumerate}
The output of step 2.4 is a realization of $\pmb{s}=(s_1,\ldots,s_T)$ from which we can recover the count matrix $N=\{n_{ij}\}$, where $n_{ij}$ is the number of times state $i$ transitions to state $j$, $i,j \in \{1,\ldots,K\}$.

To sample $\pmb{\beta}$ we introduce another set of auxiliary independent variables $m_{ij}$ such that 
     $$p(m_{ij}=m|\pmb{s},\pmb{\beta},\alpha)\propto S(n_{ij},m)(\alpha\beta_j)^m,$$
     where $S(\cdot,\cdot)$ is the unsigned Stirling number of the first kind.\footnote{By definition \citep{teh2006hierarchical}, $S(0,0)=S(1,1), S(n,0)=0$ for $n>0$ and $S(n,m)=0$ for $m>n$. Other entries can be computed as $S(n+1,m)=S(n,m-1)+nS(n,m)$.}
     Merging the infinitely many states not represented in $\pmb{s}$ into one state, the conditional distribution of $(\beta_1,\ldots,\beta_K,\sum_{k'=K+1}^{\infty}\beta_{k'})$ is 
     $$\operatorname{Dir}(m_{.1},\ldots,m_{.K},\gamma),$$
     where $m_{.k}=\sum_{k'=1}^K m_{k'k}$.

     Regarding $\boldsymbol{\pi}_k,$ ($k=1,\ldots,K$), we adopt the same strategy, merging all infinitely many states not represented into one. The conditional distribution of $(\pi_{k1},\ldots,\pi_{kK},\sum_{k'=K+1}^{\infty} \pi_{kk'})$ is then $$\operatorname{Dir}(n_{k1}+\alpha\beta_1,\ldots,n_{kK}+\alpha\beta_{K},\alpha\sum_{i=K+1}^{\infty}\beta_i).$$

Each $\phi_k$ is independent of the others conditional on $\pmb{s}, \textbf{y}$ and their prior distribution $H$, i.e.
    $$p(\pmb{\phi}|\pmb{s},\textbf{y},H)=\prod_k p(\phi_k|\pmb{s},\textbf{y},H).$$
    When $H$ is conjugate to the data distribution $F$ each $\phi_k$ can be sampled efficiently, otherwise we may resort to a Metropolis-Hastings algorithm.

We obtain MCMC samples from the posterior distributions of $\gamma$ and $\alpha$ using extensions of analogous techniques for DP.
    In particular, \cite{teh2006hierarchical} suggest using the techniques of \cite{escobar1995bayesian}.
    The concentration parameter $\gamma$ governs the distribution over the number of states, $m..=\sum_{i}\sum_{j}m_{ij}$ with $m_{ij}$ sampled at previous step,
$$p(K|\gamma,m_{..})=S(m_{..},K)\gamma^K\frac{\Gamma(\gamma)}{\Gamma(\gamma+m_{..})},$$
    setting a prior $p(\gamma)$ on $\gamma$ we get
    $$p(\gamma|K)\propto p(\gamma)p(K|\gamma,m_{..}).$$
    We may discretize the range of $\gamma$ and get a discrete approximation of the posterior.
    More attractively, sampling from the exact continuous posterior is possible when the prior $p(\gamma)$ comes from the class of mixtures of gamma distributions.
    Assuming the prior $\gamma\sim \operatorname{Gamma}(a_\gamma,b_\gamma)$
    \begin{itemize}
        \item[1] Sample an auxiliary variable $\eta|\gamma,K \sim \operatorname{Beta}(\gamma+1,m_{..})$.
        \item[2] Sample a new value for $\gamma$ from
        $$\begin{aligned}
\gamma|\eta, K & \sim \pi_\eta \operatorname{Gamma}\left(a_\gamma+K,b_\gamma-\log(\eta)\right)\\
&+(1-\pi_\eta)\operatorname{Gamma}\left(a_\gamma+K-1,b_\gamma-\log(\eta)\right),
\end{aligned}$$
        with $\pi_\eta$ such that $\frac{\pi_\eta}{1-\pi_\eta}=\frac{a_\gamma+K-1}{m_{..}(b_\gamma-\log(\eta))}$.
    \end{itemize}
    
    Sampling for $\alpha$ is similar, as it relies on a double augmentation. Let $K$ be the number of visited states, we define $\pmb{w}=(w_j)_{j=1}^K$ with $w_j \in [0,1]$ and $\pmb{q}=(q_j)_{j=1}^K$, with $q_j \in \{0,1\}$  such that
    \begin{align*}
        w_j|\alpha &\sim \operatorname{Beta}(\alpha+1,n_{j.}), \\
        q_j|\alpha &\sim \operatorname{Bernoulli}\left(\frac{n_{j.}}{\alpha}\right),
        \end{align*}
    where $n_{j.}=\sum_k n_{jk}$, i.e. the number of transitions out of state $j$.
    Then the conditional distribution $\alpha|\pmb{w}, \pmb{s}$ is given by
    $$\alpha|\pmb{w},\pmb{q} \sim \operatorname{Gamma}\left(a_\alpha+m_{..}-\sum_{j=1}^K q_j,b_\alpha-\sum_{j=1}^K\log w_j\right).$$

    \section{Adjusted Rand index}
    \label{sec:ARI}

    The adjusted Rand index \citep[ARI,][]{hubert:1985} measures the extent to which two partitions overlap: let us denote with $C^{(m)}=\{C_1^{(m)},\ldots,C_h^{(m)}\}$ the partition obtained with a generic clustering algorithm, which yields $h$ clusters on a set of $n$ elements, and with $C^{(r)}=\{C_1^{(r)},\ldots,C_{d}^{(r)}\}$  the real clustering, consisting of $d$ groups. 
Consider the following contingency table
\[
\begin{array}{c|cccc|c}
\vspace{1em}
 & C_1^{(m)} & C_2^{(m)} & \cdots & C_h^{(m)} & \sum_{j=1}^h n_{ij} \\
\hline 
C_1^{(r)} & n_{11} & n_{12} & \cdots & n_{1 h} & a_1 \\
C_2^{(r)} & n_{21} & n_{22} & \cdots & n_{2 h} & a_2 \\
\vdots & \vdots & \vdots & \ddots & \vdots & \vdots \\
C_{d}^{(r)} & n_{d 1} & n_{d 2} & \cdots & n_{d h} & a_{d} \\
\hline 
\sum_{i=1}^d n_{ij} & b_1 & b_2 & \cdots & b_h & n
\vspace{1em}
\end{array}
\]
where $n_{ij}$ is the count of elements common to clusters $C_i^{(r)}$ and $C_j^{(m)}$, $a_i=\sum_j n_{ij}$, and $b_j=\sum_i n_{ij}$.
Then,
$\text{ARI}(C^{(r)},C^{(m)})$ is calculated as \citep{ARIdef2}
\begin{equation*}
\operatorname{ARI}\left(C^{(r)}, C^{(m)}\right) :=
\frac{\sum_{j}\sum_{i}
\binom{n_{ij}}{2}
-\frac{\left[\sum_i
\binom{a_{i}}{2}
\sum_j
\binom{b_{j}}{2}
\right]}{
\binom{n}{2}
}}{\frac{1}{2}\left[\sum_i
\binom{a_{i}}{2}
+\sum_j
\binom{b_{j}}{2}
\right]-\frac{\left[\sum_i
\binom{a_{i}}{2}
\sum_j
\binom{b_{j}}{2}
\right]}{
\binom{n}{2}
}} \,.
\end{equation*}
It essentially checks, for every pair of observations, whether the two clusterings place a specific pair together or apart, and then compares the overall level of agreement. Because some agreement would occur just by chance, the ARI subtracts this expected random agreement. To compute ARI, we use the \texttt{R} function \textit{adjustedRandIndex} from the \textbf{mclust} package.

\section{Additional simulation results}
\label{sec:additionalsim}

\subsubsection*{Gaussian data}
\label{subsec:addsim_gauss}
Table \ref{tab:ari_summary} summarizes overall performance when considering Gaussian data, showing that the $k$-means and pam initializations yield the highest median ARI and lowest variability.
\begin{table}[!htbp]
\centering
\footnotesize
\caption{Median, 2.5\% and 97.5\% quantiles, standard deviation (SD) of ARI and median log-likelihood (LogLik) across initialization methods,  for data simulated from a Gaussian distribution.
Highest ARI median and lowest ARI standard deviation are highlighted in bold.}
\label{tab:ari_summary}
\begin{tabular}{lccccc}
\toprule
Method & Median & 2.5\%  & 97.5\% & SD & LogLik \\
\midrule
$k$-means   & \textbf{0.98 }& 0.38 & 1.00 & \textbf{0.19} & -8971 \\
mixtures & 0.96 & 0.30 & 1.00 & 0.22 & -9008 \\
pam      & 0.98 & 0.01 & 1.00 & 0.24 & -8969 \\
uniform  & 0.92 & 0.00 & 1.00 & 0.29 & -8895 \\
\bottomrule
\end{tabular}
\end{table}

\begin{figure}[!htbp]
    \centering
    \includegraphics[width=.8\linewidth]{ 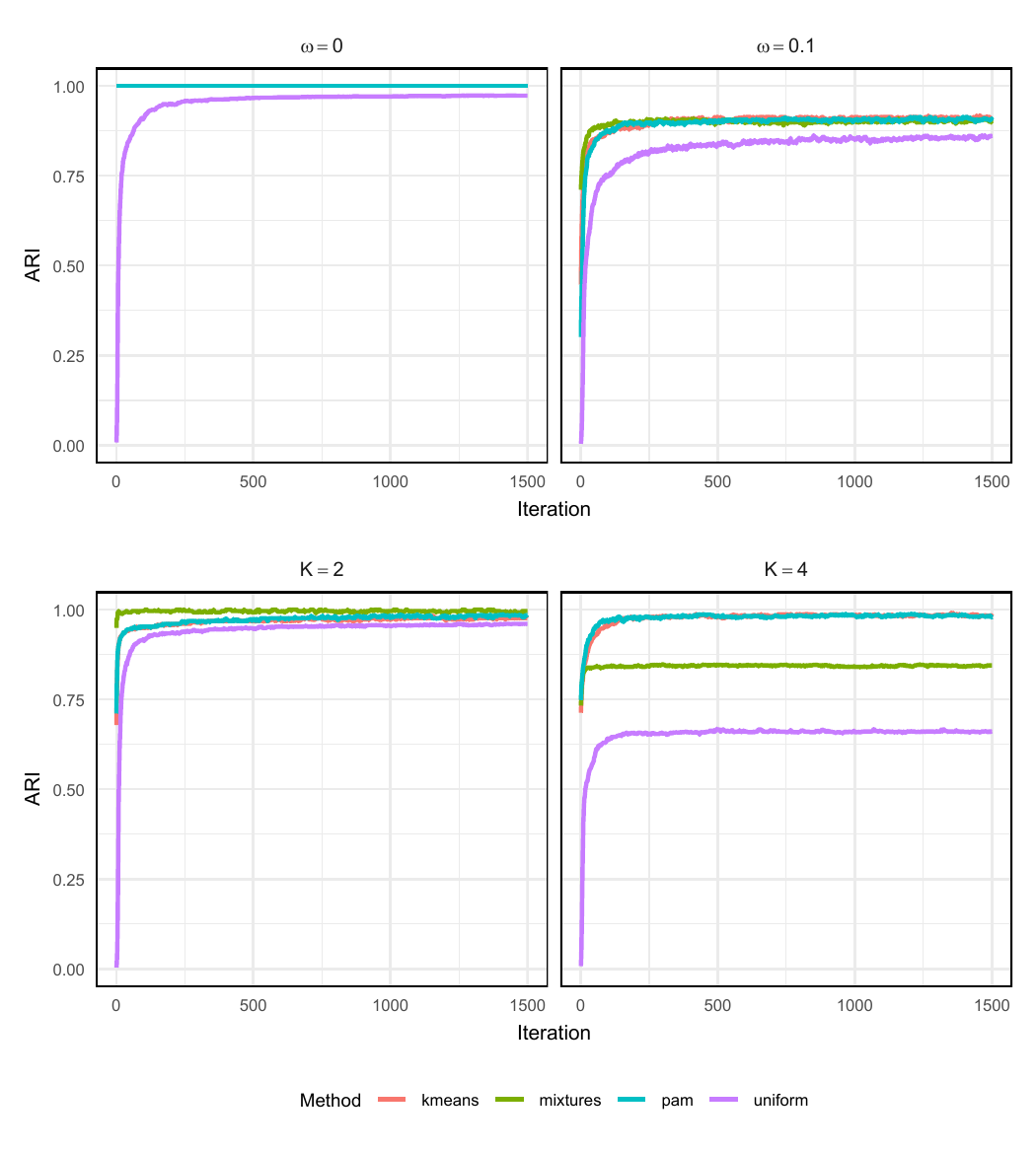}
    \caption{Top panel: median ARI across 50 seeds at each iteration, computed between true and estimated latent sequences for all initialization methods, for different $\omega$ values. Bottom panel: median ARI across 50 seeds at each iteration for different true numbers of states $K$. Results refer to data simulated from a Gaussian distribution.}
    \label{fig:varomega_varK}
\end{figure}

\noindent
Figure~\ref{fig:varomega_varK} illustrates the effect of cluster separation and the number of latent states on initialization performance for data simulated from a Gaussian distribution. The top panel reports the median ARI across iterations for different values of $\omega$. When $\omega = 0$, all methods except uniform initialization achieve perfect classification, whereas overall performance decreases when $\omega = 0.1$, with Gaussian mixtures converging quickly to the optimal solution. The bottom panel shows the median ARI across iterations for different true numbers of states $K$. $k$-means and pam again yield the best results, clearly outperforming Gaussian mixture initialization, particularly when $K = 4$.

\noindent
Table \ref{tab:ari_T_P} shows that longer sequences generally improve estimation stability, with
$k$-means and pam consistently outperforming the others.
Moreover, we observe that performance remains strong in moderate dimensions but becomes more variable as $P$ increases.
In low dimensions ($P = 5$), mixture initialization achieves the most accurate result, while in higher dimensions ($P = 20$), pam and $k$-means maintain high ARI values despite increased model complexity.
\begin{table}[!htbp]
\centering
\footnotesize
\caption{
Median, 2.5\% and 97.5\% quantiles and standard deviation (SD) of ARI across initialization methods,
for different $T$ and $P$ values, with data simulated from a Gaussian distribution.
Highest ARI median and lowest ARI standard deviation are highlighted in bold.
}
\label{tab:ari_T_P}

\begin{tabular}{@{}l|cccc@{\hskip 1.2em}|cccc@{}}
\toprule
Method & Median & 2.5\% & 97.5\% & SD & Median & 2.5\% & 97.5\% & SD \\
\midrule
\multicolumn{5}{c}{\textbf{$T = 500$}} & \multicolumn{4}{c}{\textbf{$P = 5$}} \\[3pt]
$k$-means & \textbf{0.98} & 0.35 & 1.00 & \textbf{0.20} & 0.99 & 0.46 & 1.00 & 0.13 \\
mixtures  & 0.94 & 0.00 & 1.00 & 0.25 & \textbf{1.00} & 0.65 & 1.00 & \textbf{0.11} \\
pam       & \textbf{0.98} & 0.00 & 1.00 & 0.28 & 0.98 & 0.48 & 1.00 & 0.13 \\
uniform   & 0.90 & 0.00 & 1.00 & 0.35 & 0.95 & 0.45 & 1.00 & 0.17 \\[6pt]

\multicolumn{5}{c}{\textbf{$T = 1000$}} & \multicolumn{4}{c}{\textbf{$P = 20$}} \\[3pt]
$k$-means & \textbf{0.98} & 0.42 & 1.00 & \textbf{0.18} & 0.97 & 0.33 & 1.00 & \textbf{0.24} \\
mixtures  & 0.96 & 0.34 & 1.00 & 0.19 & 0.94 & 0.00 & 1.00 & 0.28 \\
pam       & \textbf{0.98} & 0.39 & 1.00 & 0.20 & \textbf{0.98} & 0.00 & 1.00 & 0.30 \\
uniform   & 0.94 & 0.39 & 1.00 & 0.21 & 0.84 & 0.00 & 1.00 & 0.35 \\
\bottomrule
\end{tabular}
\end{table}

\noindent
Table~\ref{tab:geweke_act_T_P} reports the Geweke success rates and the 50 and 97.5\% quantiles of the ACT across different sample lengths ($T$) and dimensionalities ($P$).
Overall, $k$-means, pam, and mixtures initializations show high Geweke success rates (above 0.85) and stable convergence, while the uniform initialization performs worse, particularly for shorter series ($T = 500$).
The 97.5\% ACT quantiles remain below~2 for most configurations, indicating efficient mixing, except for the uniform initialization.
\begin{table}[!htbp]
\centering
\footnotesize
\caption{
Mean and standard deviation (SD) of Geweke success rate, and median and 97.5\% quantile of the autocorrelation time (ACT) across initialization methods,
for different $T$ and $P$ values.}
\label{tab:geweke_act_T_P}

\begin{tabular}{@{}l|cccc@{\hskip 1.2em}|cccc@{}}
\toprule
Method & Geweke & SD Geweke & ACT & 97.5\% ACT & Geweke & SD Geweke & ACT & 97.5\% ACT \\
\midrule
\multicolumn{5}{c}{\textbf{$T = 500$}} & \multicolumn{4}{c}{\textbf{$P = 5$}} \\[3pt]
$k$-means & 0.87 & 0.12 & 1.00 & 1.42 & 0.88 & 0.12 & 1.00 & 1.81 \\
mixtures  & 0.88 & 0.12 & 1.00 & 1.32 & 0.90 & 0.09 & 1.00 & 1.23 \\
pam       & 0.86 & 0.15 & 1.00 & 1.73 & 0.87 & 0.13 & 1.00 & 1.61 \\
uniform   & 0.81 & 0.17 & 1.00 & 2.18 & 0.85 & 0.16 & 1.00 & 6.01 \\[6pt]

\multicolumn{5}{c}{\textbf{$T = 1000$}} & \multicolumn{4}{c}{\textbf{$P = 20$}} \\[3pt]
$k$-means & 0.88 & 0.11 & 1.00 & 1.56 & 0.88 & 0.11 & 1.00 & 1.20 \\
mixtures  & 0.91 & 0.07 & 1.00 & 1.12 & 0.88 & 0.10 & 1.00 & 1.18 \\
pam       & 0.88 & 0.11 & 1.00 & 1.23 & 0.87 & 0.13 & 1.00 & 1.24 \\
uniform   & 0.81 & 0.17 & 1.00 & 3.58 & 0.77 & 0.16 & 1.00 & 1.20 \\
\bottomrule
\end{tabular}
\end{table}

\subsubsection*{Student-$t$ data}
\label{subsec:addsim_studt}

Table~\ref{tab:ari_studentT} reports the overall results when considering Student-$t$ data, showing that $k$-means and pam remain the most accurate and stable approaches, with the highest median ARI and lowest variability.
\begin{table}[!htbp]
\centering
\footnotesize
\caption{Median, 2.5\% and 97.5\% quantiles, standard deviation (SD) of ARI and median log-likelihood (LogLik) across initialization methods,  for data simulated from a Student-$t$ distribution.
Highest ARI median and lowest ARI standard deviation are highlighted in bold.}
\label{tab:ari_studentT}

\begin{tabular}{@{}lccccc@{}}
\toprule
Method & Median & 2.5\% & 97.5\% & SD & LogLik \\
\midrule
$k$-means   &\textbf{ 0.95} & 0.01 & 1.00 & \textbf{0.27} & -8756 \\
mixtures & 0.76 & 0.00 & 1.00 & 0.36 & -8941 \\
pam      & \textbf{0.95} & 0.00 & 1.00 & 0.31 & -8748 \\
uniform  & 0.59 & 0.00 & 1.00 & 0.34 & -9393 \\
\bottomrule
\end{tabular}

\end{table}

\begin{figure}[!htbp]
    \centering
    \includegraphics[width=.8\linewidth]{ 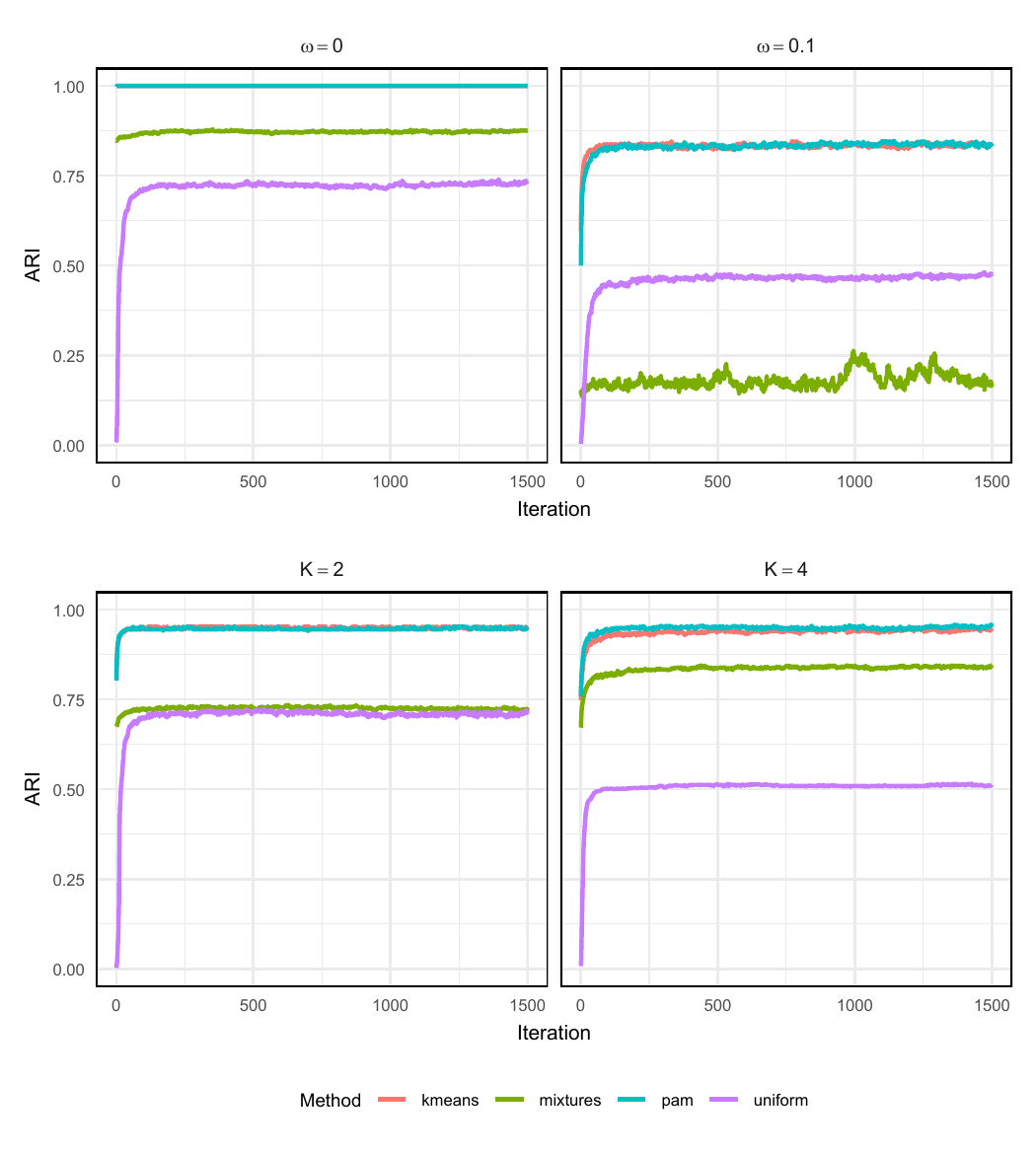}
    \caption{Top panel: median ARI across 50 seeds at each iteration, computed between true and estimated latent sequences for all initialization methods, for different $\omega$ values. Bottom panel: median ARI across 50 seeds at each iteration for different true numbers of states $K$. Results refer to data simulated from a Student-$t$ distribution.}
    \label{fig:varomega_varK_Studt}
\end{figure}

\noindent
Figure~\ref{fig:varomega_varK_Studt} summarizes the effect of cluster separation and the number of latent states on initialization performance.
The top panel shows the median ARI across iterations for different values of $\omega$: when $\omega = 0$, $k$-means and pam achieve nearly perfect classification, while mixture and uniform initializations fail to reach comparable accuracy. For $\omega = 0.1$, $k$-means and pam remain the best-performing strategies, whereas mixture initialization reaches only about 0.30 in ARI.
The bottom panel reports the median ARI across iterations for different true numbers of states $K$. pam and $k$-means again deliver the best results, outperforming Gaussian mixture initialization in both accuracy and convergence speed.

\noindent
Table~\ref{tab:ari_studentT_T_P} summarizes results for varying sequence length $T$ and number of variables $P$.
Increasing $T$ has a small effect on performance, while higher dimensionality ($P = 20$) leads to slightly lower accuracy and greater variability across initialization methods, particularly for mixtures and uniform initialization.
\begin{table}[!htbp]
\centering
\footnotesize
\caption{
Median, 2.5\% and 97.5\% quantiles and standard deviation (SD) of ARI across initialization methods,
for different $T$ and $P$ values, with data simulated from a Student-$t$ distribution.
Highest ARI median and lowest ARI standard deviation are highlighted in bold.
}
\label{tab:ari_studentT_T_P}

\begin{tabular}{@{}l|cccc@{\hskip 1.2em}|cccc@{}}
\toprule
Method & Median & 2.5\% & 97.5\% & SD & Median & 2.5\% & 97.5\% & SD \\
\midrule
\multicolumn{5}{c}{\textbf{$T = 500$}} & \multicolumn{4}{c}{\textbf{$P = 5$}} \\[3pt]
$k$-means & \textbf{0.97} & 0.01 & 1.00 & \textbf{0.28} & \textbf{0.95} & 0.44 & 1.00 & 0.16 \\
mixtures  & 0.76 & 0.00 & 1.00 & 0.37 & 0.84 & 0.61 & 1.00 & \textbf{0.12} \\
pam       & 0.96 & 0.00 & 1.00 & 0.31 & \textbf{0.95} & 0.46 & 1.00 & 0.15 \\
uniform   & 0.62 & 0.00 & 1.00 & 0.37 & 0.76 & 0.38 & 1.00 & 0.20 \\[6pt]

\multicolumn{5}{c}{\textbf{$T = 1000$}} & \multicolumn{4}{c}{\textbf{$P = 20$}} \\[3pt]
$k$-means & \textbf{0.94} & 0.01 & 1.00 & \textbf{0.26} & 0.97 & 0.00 & 1.00 & \textbf{0.34} \\
mixtures  & 0.76 & 0.00 & 0.99 & 0.36 & 0.20 & 0.00 & 0.92 & 0.39 \\
pam       & \textbf{0.94} & 0.00 & 1.00 & 0.30 & \textbf{0.98} & 0.00 & 1.00 & 0.39 \\
uniform   & 0.56 & 0.00 & 1.00 & 0.32 & 0.26 & 0.00 & 0.91 & 0.32 \\
\bottomrule
\end{tabular}
\end{table}

\noindent
Table~\ref{tab:geweke_act_T_P_studt} summarizes the Geweke success rates and the 50 and 97.5\% ACT quantiles for different time lengths ($T$) and dimensionalities ($P$) under Student-$t$ emissions.
Overall, Geweke success rates are slightly lower than in the Gaussian case.
Nevertheless, $k$-means and pam maintain relatively stable convergence, while mixtures and uniform initializations show more variability, particularly for shorter series and smaller $P$.
The 97.5 \% ACT quantiles are generally higher across all methods.
\begin{table}[!htbp]
\centering
\footnotesize
\caption{
Mean and standard deviation (SD) of Geweke success rate, and median and 97.5\% quantile of the autocorrelation time (ACT) across initialization methods,
for different $T$ and $P$ values (Student-$t$ data).}
\label{tab:geweke_act_T_P_studt}

\begin{tabular}{@{}l|cccc@{\hskip 1.2em}|cccc@{}}
\toprule
Method & Geweke & SD Geweke & ACT & 97.5\% ACT & Geweke & SD Geweke & ACT & 97.5\% ACT \\
\midrule
\multicolumn{5}{c}{\textbf{$T = 500$}} & \multicolumn{4}{c}{\textbf{$P = 5$}} \\[3pt]
$k$-means & 0.83 & 0.18 & 1.00 & 3.91 & 0.82 & 0.17 & 1.00 & 2.93 \\
mixtures  & 0.80 & 0.18 & 1.15 & 4.84 & 0.77 & 0.17 & 1.46 & 5.21 \\
pam       & 0.82 & 0.20 & 1.00 & 3.98 & 0.81 & 0.17 & 1.00 & 3.32 \\
uniform   & 0.79 & 0.17 & 1.00 & 4.53 & 0.77 & 0.19 & 1.26 & 6.75 \\[6pt]
\multicolumn{5}{c}{\textbf{$T = 1000$}} & \multicolumn{4}{c}{\textbf{$P = 20$}} \\[3pt]
$k$-means & 0.82 & 0.18 & 1.00 & 2.67 & 0.83 & 0.20 & 1.00 & 6.34 \\
mixtures  & 0.80 & 0.16 & 1.20 & 3.72 & 0.84 & 0.17 & 1.12 & 2.15 \\
pam       & 0.84 & 0.17 & 1.00 & 3.26 & 0.84 & 0.20 & 1.00 & 5.22 \\
uniform   & 0.80 & 0.18 & 1.05 & 4.64 & 0.82 & 0.16 & 1.00 & 2.18 \\
\bottomrule
\end{tabular}
\end{table}

\clearpage
\section{Empirical analysis on industrial production index data excluding the COVID-19 period}
\label{sec:addemp}

\begin{figure}[!htbp]
    \centering
    \includegraphics[width=.8\linewidth]{ 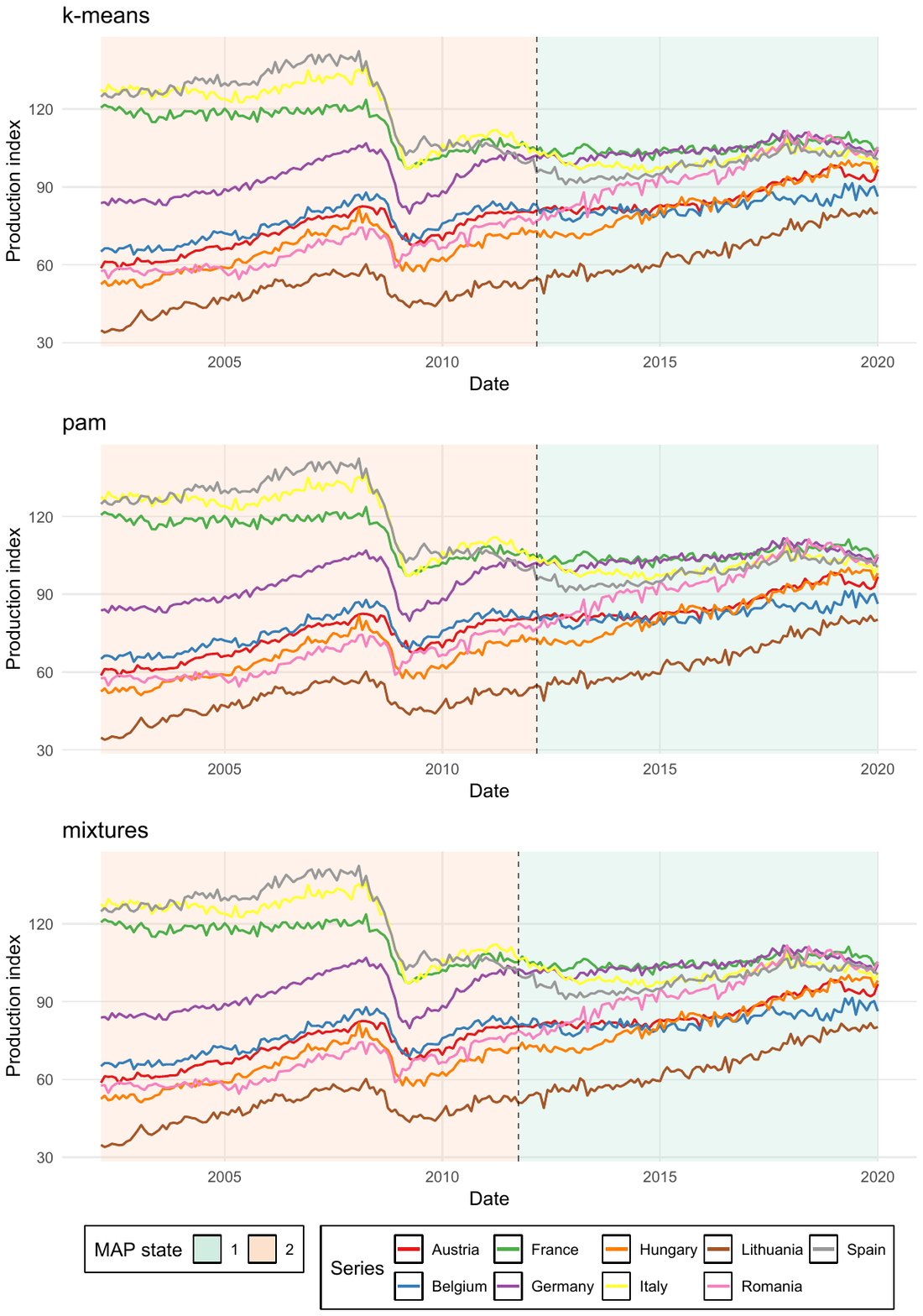}
    \caption{Regime classification over time for each initialization method, and production index time-series of each country.}
    \label{fig:stateclass_nocovid}
\end{figure}

We replicate the analysis of Section~5.1 using data up to January~2020, just before the COVID-19 outbreak.
Across initialization methods, the iHMMs consistently identify two latent states ($K=2$), corresponding to low- and high-volatility regimes.
The estimated covariance traces in Table~\ref{tab:stateCond_var_noCOVID} confirm this pattern, with the high-volatility state exhibiting roughly twice the overall variability of the low-volatility one.
State-conditional mean levels, reported in Table~\ref{tab:stateCond_means_noCOVID}, are also highly consistent across initialization methods.
Figure~\ref{fig:stateclass_nocovid} shows the temporal regime classification based on the maximum a posteriori of estimated state probabilities.
Change points are detected in February~2012 for the $k$-means and pam initializations, and in September~2011 for the mixture initialization.

\begin{table}[!htbp]
\centering
\caption{Trace of the covariance matrices for each inferred state and initialization method.}
\label{tab:stateCond_var_noCOVID}
\begin{tabular}{lcc}
\toprule
Method & State 1 & State 2 \\
\midrule
$k$-means & 337.00 & 652.00 \\
pam       & 340.00 & 647.00 \\
mixtures  & 357.00 & 622.00 \\
\bottomrule
\end{tabular}
\end{table}

\begin{table}[!htbp]
\centering
\caption{State-conditional mean values by initialization method.}
\label{tab:stateCond_means_noCOVID}
\footnotesize
\begin{tabular}{lrrrrrrrrr}
\toprule
 & Austria & Belgium & France & Germany & Hungary & Italy & Lithuania & Romania & Spain \\
\midrule
\multicolumn{10}{l}{\textbf{$k$-means}} \\[2pt]
State 1 & 86.63 & 82.79 & 105.24 & 104.50 & 84.45 & 100.62 & 66.18 & 95.63 & 98.63 \\
State 2 & 71.74 & 75.28 & 113.68 & 92.98  & 64.38 & 120.65 & 48.33 & 65.16 & 122.87 \\
\addlinespace[4pt]
\multicolumn{10}{l}{\textbf{pam}} \\[2pt]
State 1 & 86.53 & 82.78 & 105.24 & 104.46 & 84.32 & 100.68 & 66.00 & 95.44 & 98.65 \\
State 2 & 71.59 & 75.19 & 113.78 & 92.84  & 64.25 & 120.87 & 48.24 & 64.97 & 123.15 \\
\addlinespace[4pt]
\multicolumn{10}{l}{\textbf{mixtures}} \\[2pt]
State 1 & 86.28 & 82.73 & 105.26 & 104.34 & 83.83 & 100.87 & 65.45 & 94.65 & 98.68 \\
State 2 & 71.32 & 74.99 & 114.03 & 92.56  & 64.01 & 121.35 & 48.12 & 64.63 & 123.92 \\
\bottomrule
\end{tabular}
\end{table}

\section{An application to the Old Faithful geyser dataset}
\label{sec:old}

This application analyzes the well-known \textit{Old Faithful} geyser data, available in the \textbf{datasets} \texttt{R} package. 
It consists of $T = 272$ bivariate observations: the eruption duration and the waiting time to the next eruption, both measured in minutes.

Across the different initialization strategies, the iHMM exhibits broadly consistent convergence behavior. 
Under the $k$-means initialization, seven of the ten chains estimate a median number of states equal to 2.
Among these, five chains converge according to the Geweke and ACT criteria introduced earlier.
Results obtained using the pam initialization are comparable: nine chains select two regimes, six of which meet the convergence criteria. 
When using the Gaussian mixture initialization, six chains favor three regimes, while the remaining chains estimate two states. The only two convergent chains belong to this latter group.
For these three initialization methods, the Gelman–Rubin diagnostic confirms excellent mixing, with median \(\hat{R}\) values close to one across all parameters in the convergent chains.
In contrast, the uniform initialization performs poorly: all chains display long ACTs, and achieve low Geweke success rates.
%

\begin{figure}
    \centering
    \includegraphics[width=.7\linewidth]{ 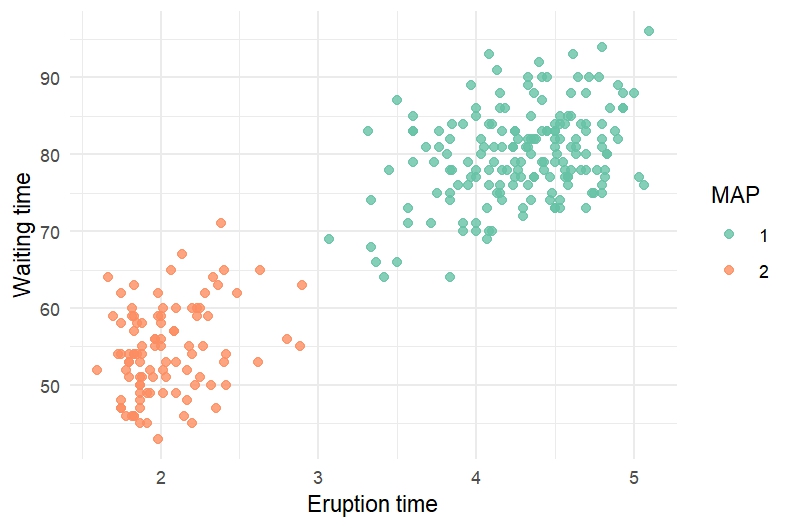}
\caption{State classification for the \textit{Old Faithful} geyser data based on the \textit{maximum a posteriori} (MAP) probabilities obtained from the iHMM with \textit{k}-means initialization.}
    \label{fig:old_faith}
\end{figure}

Figure~\ref{fig:old_faith} shows the classification of the observations based on the MAP state probabilities obtained under the \textit{k}-means initialization. 
The results for the convergent chains using alternative initialization methods are identical.
In particular, we notice two regimes that identify short eruptions followed by brief waiting times and long eruptions followed by long waiting times. These results are in line with the results obtained from prior analysis using finite HMMs and mixture models \citep{maruotti2021initialization}.

\end{document}